\documentclass[twocolumn]{aastex631}

\usepackage{amsmath}

\newcommand{\eg}{e.g., }
\newcommand{\ie}{i.e., }
\newcommand{\Msun}{M_{\odot}}

\newcommand{\Rcsm}{r_{\rm CSM}}
\newcommand{\Menv}{M_{\rm env}}
\newcommand{\XH}{X({\rm H})_{\rm env}}

\newcommand{\Lp}{L_{\rm pk}}

\newcommand{\Tp}{T_{\rm pk}}
\newcommand{\Lpl}{L_{\rm pl}}
\newcommand{\taupl}{\tau_{\rm pl}}
\newcommand{\taucsm}{\tau_{\rm CSM}}

\newcommand{\Nifs}{$^{56}$Ni}
\newcommand{\Mni}{M{\rm (^{56}Ni)}}
\newcommand{\Mms}{M_{\rm ZAMS}}
\newcommand{\Mej}{M_{\rm ej}}
\newcommand{\Mdot}{\dot{M}}
\newcommand{\vw}{v_{\rm wind}}
\newcommand{\vi}{v_{\rm 0}}
\newcommand{\vf}{v_{\rm \infty}}

\newcommand{\Eexp}{E_{\rm exp}}
\newcommand{\Myr}{\Msun~{\rm yr}^{-1}}
\newcommand{\rph}{r_{\rm ph}}
\newcommand{\Tph}{T_{\rm ph}}

\newcommand{\Tcol}{T_{\rm col}}
\newcommand{\rrec}{r_{\rm rec}}
\newcommand{\Rout}{r_{\rm out}}

\newcommand{\tPT}{t_{\rm PT}}
\def\gsim{\mathrel{\rlap{\lower 4pt \hbox{\hskip 1pt $\sim$}}\raise 1pt
\hbox {$>$}}}
\def\lsim{\mathrel{\rlap{\lower 4pt \hbox{\hskip 1pt $\sim$}}\raise 1pt
\hbox {$<$}}}

\begin{document}

\title{A Robust Light-Curve Diagnostic for Electron-Capture Supernovae and Low-Mass Fe-Core-Collapse Supernovae}

\author[0009-0008-0189-863X]{Masato Sato}
\altaffiliation{e-mail: masato.sato@grad.nao.ac.jp}
\affiliation{Astronomical Science Program, Graduate Institute for Advanced Studies, SOKENDAI 2-21-1 Osawa, Mitaka, Tokyo 181-8588, Japan}
\affiliation{National Astronomical Observatory of  Japan, National Institutes of Natural Sciences 2-21-1 Osawa, Mitaka, Tokyo 181-8588, Japan}

\author[0000-0001-8537-3153]{Nozomu Tominaga}
\affiliation{National Astronomical Observatory of  Japan, National Institutes of Natural Sciences 2-21-1 Osawa, Mitaka, Tokyo 181-8588, Japan}
\affiliation{Astronomical Science Program, Graduate Institute for Advanced Studies, SOKENDAI 2-21-1 Osawa, Mitaka, Tokyo 181-8588, Japan}
\affiliation{Department of Physics, Faculty of Science and Engineering, Konan University, 8-9-1 Okamoto, Kobe, Hyogo 658-8501, Japan}

\author[0000-0002-5726-538X]{Sergei I. Blinnikov}
\affiliation{NRC Kurchatov Institute 123182 Moscow, Russia}
\affiliation{Dukhov Automatics Research Institute (VNIIA) 127055 Moscow, Russia}

\author[0000-0002-0564-1101]{Marat Sh. Potashov}
\affiliation{NRC Kurchatov Institute 123182 Moscow, Russia}
\affiliation{Keldysh Institute of Applied Mathematics RAS, 4 Miusskaya Square, 125047 Moscow, Russia}

\author[0000-0003-1169-1954]{Takashi J. Moriya}
\affiliation{National Astronomical Observatory of  Japan, National Institutes of Natural Sciences 2-21-1 Osawa, Mitaka, Tokyo 181-8588, Japan}
\affiliation{Astronomical Science Program, Graduate Institute for Advanced Studies, SOKENDAI 2-21-1 Osawa, Mitaka, Tokyo 181-8588, Japan}
\affiliation{School of Physics and Astronomy, Monash University, Clayton, Victoria 3800, Australia}

\author[0000-0002-1125-9187]{Daichi Hiramatsu}
\affiliation{Center for Astrophysics \textbar~Harvard \& Smithsonian 60 Garden Street, Cambridge, MA 02138-1516, USA}
\affiliation{The NSF AI Institute for Artificial Intelligence and Fundamental Interactions, USA}


\begin{abstract}
Core-collapse supernovae (CCSNe) are the terminal explosions of massive stars.
While most massive stars explode as iron-core-collapse supernovae (FeCCSNe), slightly less massive stars explode as electron-capture supernovae (ECSNe), 
shaping the low-mass end of CCSNe. 
ECSNe was proposed $\sim 40$ years ago and first-principles simulations also predict their successful explosions. 
Observational identification and investigation of ECSNe are important for the completion of stellar evolution theory.
To date, only one promising candidate has been proposed, SN~2018zd, other than the historical progenitor of the Crab Nebula, SN~1054. 
We present representative synthetic light curves of low-mass FeCCSNe and ECSNe exploding with energies in circumstellar media (CSM) estimated with theoretically or observationally plausible methods.
The plateaus of the ECSNe are shorter, brighter, and bluer than those of the FeCCSNe.
To investigate the robustness of their intrinsic differences, 
we adopted  various explosion energies and CSM.
Although they may have similar bolometric light-curve plateaus, ECSNe are bluer than FeCCSNe in the absence of strong CSM interaction, 
illustrating that 
multicolor observations are essential to identify ECSNe.
This provides a robust indicator of ECSNe
because the bluer plateaus stem from the low-density envelopes of their super-asymptotic-giant-branch progenitors.
Furthermore, we propose a distance-independent method to identify ECSNe:
$(g-r)_{\tPT/2} < 0.008 \times \tPT - 0.4$, 
i.e., blue $g-r$ at the middle of the plateau $(g-r)_{\tPT/2}$, 
where $\tPT$ is the transition epoch from plateau to tail. 
Using this method, we identified SN~2018zd as an ECSN, which we believe to be the first ECSN identified with modern observing techniques.
\end{abstract}



\section{Introduction \label{sec:intro}}
 Core-collapse supernovae (CCSNe) are the terminal explosions of massive stars, triggered by the collapse of their cores. Collapsing cores are divided into two mass-dependent types: Iron (Fe) cores, for which progenitors are massive enough to ignite static Si burning, 
collapse because of photo-disintegration and the stars explode as Fe-core-collapse supernovae (FeCCSNe, \citealt{Colgate1966-cq, Bethe1985-kh, Woosley2002-sx}).
Oxygen-neon-magnesium (ONeMg) cores, for which progenitors are not massive enough to ignite static O burning, are supported by electron degeneracy pressure. 
Degenerate ONeMg cores collapse owing to electron capture by Mg and Ne, and the stars explode as electron-capture supernovae (ECSNe, \citealt{Miyaji1980-mf, Nomoto1982-zj, Nomoto1984-au, Nomoto1987-fr}).
 The progenitors of ECSNe are likely to be less massive than those of FeCCSNe, shaping the low-mass end of the CCSN-progenitor distribution.
 However, the stellar mass range that forms ECSNe is not clear and may be narrow or even non-existent for solar metallicity \citep{Poelarends2008-hq, Langer2011-zv, Doherty2015-wv, Limongi2023-db}.

 If these progenitors retain a massive hydrogen-rich (H-rich) envelope, red supergiants (RSGs) result in FeCCSNe, whereas super-asymptotic-giant-branch (SAGB) stars result in ECSNe. Both types of explosion are observed as H-rich type II supernovae (SNe~II).
 After an initial shock-breakout flash, a plateau phase is powered by the shock-heated H-rich envelope.
 The duration and brightness  of a plateau infer the explosion energy, pre-supernova radius, and envelope mass of the star \citep{Litvinova1985-ty, Popov1993-cu, Eastman1994-hs, Kasen2009-qk}. 
 A tail phase following the plateau phase is powered by the radioactive decay of \(^{56}\)Co, the daughter isotope of \(^{56}\)Ni synthesized by explosive nucleosynthesis. 
 Thus, the brightness of the tail is proportional to the mass of the ejected \(^{56}\)Ni [$\Mni$]. 
 
 Observations of the light curves and spectra of SNe~II reveal diversity in properties such as the explosion energy ($\sim 10^{50}-10^{51}$ erg) and $\Mni$ ($\sim 10^{-3}-10^{-1}~\Msun$) \citep{Martinez2022-ab, Martinez2022-lv, Martinez2022-ks, Anderson2019-zb}.
 Furthermore, recent observations of early light curves and spectra indicate the presence of dense circumstellar media (CSM) around progenitors, with estimated mass-loss rates ranging from $\Mdot = 10^{-4}$ to $10^{-2}$ $\Myr$ \citep{Yaron2017-lq, Morozova2017-oj, Forster2018-ox, Bruch2021-mg, Bruch2023-nl, Subrayan2023-md}.

The explosion mechanisms of FeCCSNe are largely investigated with numerical simulations. 
It has been shown that the FeCCSN of 15$\Msun$ does not explode when assuming spherical symmetry \citep{Sumiyoshi2005-ri}. 
Recent three-dimensional (3D) simulations demonstrated that multi-dimensional effects like convection are key to  successful explosions \citep{Takiwaki2012-jo}, although the explosion energy and amount of \(^{56}\)Ni are slightly lower than those of observed SNe~II \citep{Burrows2019-zb, Sieverding2020-xs}. 

3D simulations of low-mass FeCCSNe of stars with zero-age main-sequence masses $\Mms$ of $9-11\Msun$ [s9.0--s11.0 models in \citet{Sukhbold2016}] were performed by \citet{Burrows2019-zb, Burrows2021-ts}, who found successful explosions with energies of \(\sim10^{50}\) erg.
Based on the 3D simulations \citet{Wang2023-gv} showed that such supernovae have \Nifs\ yields of $\Mni\sim 10^{-3}-10^{-2}\Msun$.
The progenitor radii range $2.9 \times 10^{13}-4.4 \times 10^{13}$ cm.
Their typical bolometric light curves have also been calculated by \citet{Sukhbold2016},
showing plateaus of $\sim-15$ to $-18$ mag, tails of $\sim-12.5$ to $-15.5$ mag at the onset, and a drop of $\sim2.5$ mag from plateau to tail. 

On the other hand, ECSN explosions have been successfully simulated with the assumption of spherical symmetry owing to the low-density H/He envelopes of their progenitors \citep{Kitaura2006-ia}. 
First-principles simulations revealed low explosion energies (\(1.5\times 10^{50}\) erg) and small amounts of \(^{56}\)Ni (\(0.002-0.004\) $\Msun$) \citep{Kitaura2006-ia, Janka2008-ai, Wanajo2009-yo}. 
The elemental abundance ratios of ECSNe are consistent with the Crab Nebula, a remnant of SN~1054 \citep{Davidson1982-xe, Nomoto1982-zj, Wanajo2009-yo}.
Moreover, \citet{Tominaga2013, Moriya2014-oh} calculated ECSN light curves and found that they exhibit a plateau with a bolometric brightness of $-15$ to $-16$ mag (as bright as that of SNe~II of RSG progenitors), a faint tail of $\sim -11$ mag at the onset, and a large drop of $\sim 4$ mag from plateau to tail. These calculations were performed using SAGB progenitor models with radii ranging $6.5 \times 10^{13}-7.3 \times 10^{13}$ cm.

Despite successful explosions in first-principles simulations, ECSN observations remain elusive.
Recently, \citet{Hiramatsu2021-er} proposed that SN~2018zd is an ECSN based on observational features and indications such as the light curve, explosion energy, dense CSM, nucleosynthesis, and pre-supernova photometry. 
However, \citet{Callis2021-zu} controversially proposed that it is a low-mass FeCCSN because of a large $\Mni$ 
using a larger distance estimation than \citet{Hiramatsu2021-er}.
Therefore, a distance-independent diagnostic method is required to firmly identify ECSNe.

The theoretical light curves of low-mass FeCCSNe and ECSNe have been calculated by \citet{Kozyreva2021-ci}.
They showed that ECSNe have a bluer color until the middle of the plateau in the absence of factors affecting plateau color, including explosion-energy variation and CSM.
Meanwhile, observations of SNe~II \citep{Yaron2017-lq, Morozova2017-oj, Forster2018-ox, Bruch2021-mg, Bruch2023-nl} indicate the existence of dense CSM corresponding mass-loss rates of $\Mdot=10^{-4}-10^{-2}$~$\Myr$ in the vicinity of the progenitors, and observations of RSG stars in the Large Magellanic Cloud \citep{Goldman2016-hb} and Small Magellanic Cloud \citep{Yang2023-dn} indicate mass-loss rates of $\Mdot=10^{-6}$ to $10^{-5}$ $\Myr$ distributed up to $10^{-3}$ $\Myr$. Also, the mass-loss rate of SAGB stars is expected to be $10^{-4}$ $\Myr$~from theoretical calculation \citep{Poelarends2008-hq}.
Additionally, low-mass SNe~II ($\Mms \leq 12.0~\Msun$) have an observed explosion-energy diversity of $\Eexp \sim 10^{50}-10^{51}$ erg \citep{Martinez2022-lv}, while the low explosion energy theoretically estimated for low-mass FeCCSNe and ECSNe ($\sim10^{50}$~erg, \citealt{Burrows2019-zb, Burrows2021-ts, Kitaura2006-ia, Janka2008-ai}). 

Therefore, we calculated and investigated the multicolor light curves of low-mass FeCCSNe and ECSNe as follows.
First, we adopted the physical quantities estimated in previous studies with theoretically or observationally plausible methods such as the explosion energy from explosion simulation and mass-loss rates from progenitor observation or calculation 
to derive representative light curves. 
Then, we adopt wide ranges of physical quantities to robustly discriminate 
between ECSNe and low-mass FeCCSNe. 

This paper is structured as follows.
We introduce the models and methods in Section~\ref{sec:method}.
We show the resulting light curves in Section~\ref{sec:results}: the representative light curves in Section~\ref{subsec:typicalLC} and light curves with a wide range of physical quantities in Section~\ref{subsec:allLC}. We discuss their robust characteristics and propose an ECSN diagnostic method in Section~\ref{sec:discussion}. Finally, we present our conclusions in Section~\ref{sec:conclusion}. 

\section{Model \& Method \label{sec:method}}
We adopted SAGB and RSG progenitor models to calculate light curves for ECSNe and low-mass FeCCSNe, respectively. 
On top of the both progenitor models, 
we attached a CSM structure following \citet{Moriya2018-ve}, with density
\begin{equation}
\rho_{CSM}(r) = \frac{\Mdot}{(4\pi\vw)r^{-2}},
\label{eq:CSMdensity}
\end{equation}
up to $r=\Rout$, where $r$ is the distance from the center of the star, $\vw$ is the wind velocity, and $\Rout$ is the CSM radius. For the wind velocity, acceleration was considered using a simple $\beta$ velocity law:
\begin{equation}
\vw(r) = \vi + (\vf-\vi)(1-\frac{R}{r})^\beta ,
\label{eq:windvel}
\end{equation}
where $\vi$ is the initial velocity of the wind, $\vf$ is the terminal velocity of the wind, and $R$ is the progenitor radius.

We adopted the SAGB progenitor models from \citet{Tominaga2013}. 
Since SAGB stars experience uncertain mass loss and third dredge-up from thermal pulses,
we adopted various envelope masses $\Menv$ and H abundances in the envelope $\XH$. 
Nucleosynthesis was adopted from \citet{Wanajo2009-yo}. 
We adopted an inner mass cut at $1.364$ or $1.370~\Msun$ to reduce the calculation cost, although it is estimated as $1.362~\Msun$ in the first principles simulation \citep{Kitaura2006-ia}. 
This small difference does not substantially affect the light curve properties. 
An explosion energy of $\Eexp = 1.5\times10^{50}$~erg from the first-principles simulation \citep{Kitaura2006-ia} and mass-loss rate of $\Mdot = 10^{-4}$ $\Myr$ from the calculation of \citet{Poelarends2008-hq} are adopted 
for the representative light curves,
as listed in Table~\ref{table:typicalparams_ec}. 
We also calculated models with wide parameter ranges,
as listed in Table~\ref{table:calcparams_EC}. 

\begin{table*}[t]
 \caption{Parameter sets for the representative ECSN light curves}
 \label{table:typicalparams_ec}
 \centering
  \begin{tabular}{ccccccccccc}
   \hline
   \(\Menv\) & \(\XH\) & $R$ & Mass cut & \(\Eexp\) & $\Mni$ & \(\Mdot\) & $\Rcsm$ & \(\vi\) & \(\vf\) & \(\beta\) \\ \relax
   [$\Msun$] & & [$10^{13}$ cm] & [$\Msun$] & [\(10^{50}\) ergs] & [$\Msun$] & [$\Myr$] & [cm] & [km/s] & [km/s] & \\
   \hline
   3.0 & 0.2 & 7.0 & 1.370 & 1.5 & 0.002 & \(10^{-4}\) & \(10^{15}\) & 1 & 10 & 3 \\
   3.0 & 0.7 & 7.1 & 1.370 & 1.5 & 0.002 & \(10^{-4}\) & \(10^{15}\) & 1 & 10 & 3 \\
   4.7 & 0.7 & 7.2 & 1.370 & 1.5 & 0.002 & \(10^{-4}\) & \(10^{15}\) & 1 & 10 & 3 \\
   \hline
  \end{tabular}
\tablecomments{
\(\Menv\) and \(\XH\) are the envelope mass and hydrogen abundance of the progenitor, $R$ is the progenitor radius, \(\Eexp\) is the explosion energy, \(\Mdot\) is the mass-loss rate, $\Rcsm$ is the CSM radius, \(\vi\) is the initial velocity of the wind, \(\vf\) is the terminal velocity of the wind, and \(\beta\) is the acceleration parameter of the wind.
}
\end{table*}

\begin{table*}[t]
 \caption{Parameter sets for the ECSN light curves with wide physical quantity ranges}
 \label{table:calcparams_EC}
 \centering
  \begin{tabular}{ccccccccccc}
   \hline
   \(\Menv\) & \(\XH\) & $R$ & Mass cut & \(\Eexp\) & $\Mni$  & \(\Mdot\) & $\Rcsm$ & \(\vi\) & \(\vf\) & \(\beta\) \\ \relax
   [$\Msun$] & & [$10^{13}$ cm] & [$\Msun$] & [\(10^{50}\) ergs] & [$\Msun$] & [$\Myr$] & [cm] & [km/s] & [km/s] & \\
   \hline
   2.0 & 0.7 & 6.5 & 1.364/1.370 & $0.6-14.2$ & $0.002$ & \(10^{-6}-10^{-2}\) & \(10^{14}-10^{16}\) & $1-5$ & 10 & $1-5$ \\
   3.0 & 0.2 & 7.0 & 1.364/1.370 & $0.6-12.6$ & $0.002-0.003$ & \(10^{-6}-10^{-2}\) & \(10^{14}-10^{16}\) & $1-5$ & 10 & $1-5$ \\
   3.0 & 0.5 & 7.0 & 1.364/1.370 & $0.6-10.4$ & $0.002-0.003$ & \(10^{-4}-10^{-2}\) & \(10^{14}-10^{16}\) & $1-5$ & 10 & $1-5$ \\
   3.0 & 0.7 & 7.1 & 1.364/1.370 & $0.6-10.5$ & $0.002-0.003$ & \(10^{-6}-10^{-2}\) & \(10^{14}-10^{16}\) & $1-5$ & 10 & $1-5$ \\
   4.7 & 0.5 & 7.3 & 1.364/1.370 & $0.6-13.7$ & $0.002$ & \(10^{-6}-10^{-2}\) & \(10^{14}-10^{16}\) & $1-5$ & 10 & $1-5$ \\
   4.7 & 0.7 & 7.2 & 1.364/1.370 & $0.6-13.7$ & $0.002-0.003$ & \(10^{-6}-10^{-2}\) & \(10^{14}-10^{16}\) & $1-5$ & 10 & $1-5$ \\
   \hline
  \end{tabular}
\end{table*}

For the representative light curves, we adopted the RSG progenitor models of s9.0, s10.0, and s11.0 ($\Mms=9.0,~10.0,$ and $11.0~\Msun$ respectively) in \citet{Sukhbold2016}, explosion energies of $\Eexp = 0.9$, $1.5$, and $1.5 \times 10^{50}$ erg, for each progenitor respectively, from \citet{Burrows2021-ts}, and a mass-loss rate of $\Mdot = 10^{-6}$ $\Myr$ \citep{Goldman2016-hb, Yang2023-dn}.
The adopted parameters are listed in Table~\ref{table:typicalparams_Fe}.
We also calculated models with wide parameter ranges,
as listed in Table~\ref{table:calcparams_Fe}.
In both calculations, 
the hydrogen abundance in the progenitor envelope is approximately $0.6$, comparable to the $\XH=0.7$ models for SAGB stars.
We adopted mass cuts for $\Mms=9.0$, $10.0$, and $11.0~\Msun$ progenitors and $\Mni$ for $\Mms=9.0$, $9.25$, $9.5$, and $11.0~\Msun$ progenitors from the results of 3D simulation \citep{Burrows2021-ts, Wang2023-gv}, otherwise from the results of spherical simulation \citep{Sukhbold2016}. We placed \Nifs~at the bottom of the ejecta. 

We focus on the plateau in this study, the duration of which is extended by heating from radioactive decay \citep{Kasen2009-qk} and is dependent on the \Nifs\ yield, that is, the mass cut and explosion energy. 
However, this extension is negligible for ECSNe and low-mass FeCCSNe ($\lesssim 4\%$ according to equation (10) in \citealt{Kasen2009-qk}) 
as \Nifs\ yields are small ($\sim 10^{-3}-10^{-2}~\Msun$, \citealt{Kitaura2006-ia, Janka2008-ai, Wanajo2009-yo, Wang2023-gv}).
We did not consider \Nifs~mixing because it has little impact on plateau duration and luminosity \citep{Nakar2016-xe, Moriya2015-rl, Kozyreva2019-cw}.

\begin{figure}[ht!]
\centering
\includegraphics[width=80truemm]{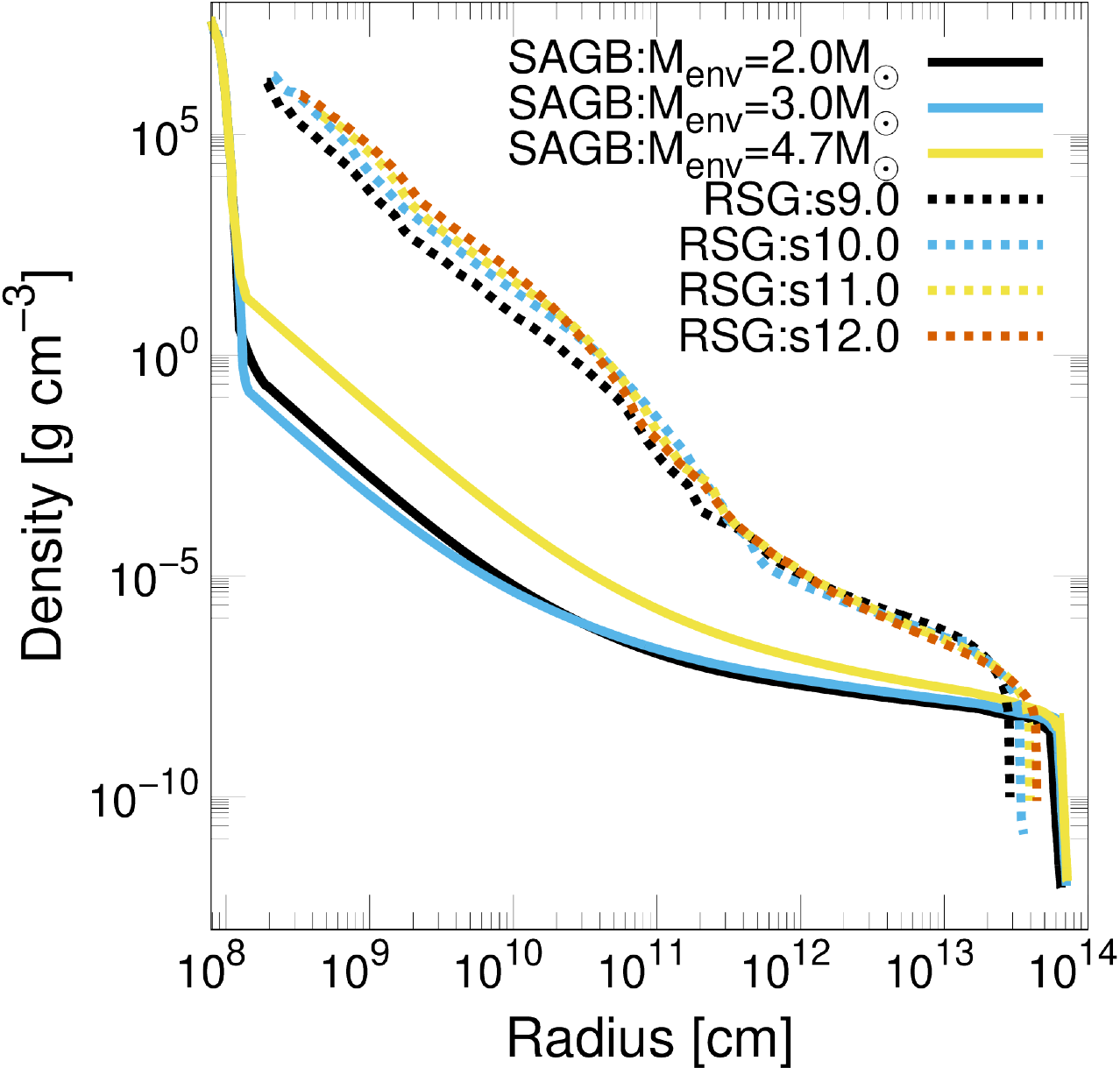}
\caption{Density profiles of select SAGB and RSG models.
\label{fig:prog_noCSM}}
\end{figure}

Examples of progenitor density profiles as a function of radius are shown in Figure~\ref{fig:prog_noCSM}. The progenitor radii are listed in Tables~\ref{table:typicalparams_ec} to \ref{table:calcparams_Fe}. SAGB stars have an extended and lower-density envelope compared to those of RSGs. 

\begin{table*}[t]
 \caption{Parameter sets for the representative light curves of low-mass FeCCSNe}
 \label{table:typicalparams_Fe}
 \centering
  \begin{tabular}{cccccccccc}
   \hline
   Progenitor model & $R$ & Mass cut & \(\Eexp\) & $\Mni$  & \(\Mdot\) & $\Rcsm$ & \(\vi\) & \(\vf\) & \(\beta\) \\
   & [$10^{13}$ cm] & [$\Msun$] & [10\(^{50}\) erg] & [$\Msun$] & [$\Myr$] & [cm] & [km/s] & [km/s] & \\
   \hline
   s9.0 & 2.9 & 1.350 & 0.9 & 0.007 & \(10^{-6}\) & \(10^{15}\) & 1 & 10 & 3 \\
   s10.0 & 3.6 & 1.490 & 1.5 & 0.03 & \(10^{-6}\) & \(10^{15}\) & 1 & 10 & 3 \\
   s11.0 & 4.0 & 1.510 & 1.5 & 0.03 & \(10^{-6}\) & \(10^{15}\) & 1 & 10 & 3 \\
   \hline
  \end{tabular}
\end{table*}

\begin{table*}[t]
 \caption{Parameter sets for the light curves of low-mass FeCCSNe with wide physical quantity ranges}
 \label{table:calcparams_Fe}
 \centering
  \begin{tabular}{cccccccccc}
   \hline
   Progenitor model & $R$ & Mass cut & \(\Eexp\) & $\Mni$  & \(\Mdot\) & $\Rcsm$ & \(\vi\) & \(\vf\) & \(\beta\) \\
   & [$10^{13}$ cm] & [$\Msun$] & [10\(^{50}\) erg] & [$\Msun$] & [$\Myr$] & [cm] & [km/s] & [km/s] & \\
   \hline
   s9.0 & 2.9 & 1.350 & $0.5-13.7$ & 0.007 & \(10^{-6}-10^{-2}\) & \(10^{14}-10^{16}\) & $1-5$ & 10 & $1-5$ \\
   s9.25 & 2.8 & 1.390 & $0.6-10.0$ & 0.01 & \(10^{-6}-10^{-2}\) & \(10^{14}-10^{16}\) & $1-5$ & 10 & $1-5$ \\
   s9.5 & 2.9 & 1.410 & $0.9-10.0$ & 0.02 & \(10^{-6}-10^{-2}\) & \(10^{14}-10^{16}\) & $1-5$ & 10 & $1-5$ \\
   s9.75 & 3.1 & 1.450 & $0.7-9.8$ & 0.01 & \(10^{-6}-10^{-2}\) & \(10^{14}-10^{16}\) & $1-5$ & 10 & $1-5$ \\
   s10.0 & 3.6 & 1.490 & $1.2-13.1$ & 0.03 & \(10^{-6}-10^{-2}\) & \(10^{14}-10^{16}\) & $1-5$ & 10 & $1-5$ \\
   s10.25 & 3.8 & 1.470 & $1.9-13.1$ & 0.03 & \(10^{-6}-10^{-2}\) & \(10^{14}-10^{16}\) & $1-5$ & 10 & $1-5$ \\
   s10.5 & 3.8 & 1.480 & $2.0-13.3$ & 0.02 & \(10^{-6}-10^{-2}\) & \(10^{14}-10^{16}\) & $1-5$ & 10 & $1-5$ \\
   s10.75 & 3.9 & 1.470 & $1.7-12.8$ & 0.03 & \(10^{-6}-10^{-2}\) & \(10^{14}-10^{16}\) & $1-5$ & 10 & $1-5$ \\
   s11.0 & 4.0 & 1.510 & $1.4-13.3$ & 0.03 & \(10^{-6}-10^{-2}\) & \(10^{14}-10^{16}\) & $1-5$ & 10 & $1-5$ \\
   s11.25 & 4.0 & 1.530 & $2.0-13.3$ & 0.02 & \(10^{-6}-10^{-2}\) & \(10^{14}-10^{16}\) & $1-5$ & 10 & $1-5$ \\
   s11.5 & 4.0 & 1.500 & $1.7-12.9$ & 0.02 & \(10^{-6}-10^{-2}\) & \(10^{14}-10^{16}\) & $1-5$ & 10 & $1-5$ \\
   s11.75 & 4.3 & 1.590 & $2.0-13.2$ & 0.02 & \(10^{-6}-10^{-2}\) & \(10^{14}-10^{16}\) & $1-5$ & 10 & $1-5$ \\
   s12.0 & 4.4 & 1.530 & $1.5-12.6$ & 0.03 & \(10^{-6}-10^{-2}\) & \(10^{14}-10^{16}\) & $1-5$ & 10 & $1-5$ \\
   \hline
  \end{tabular}
\end{table*}

We adopted the one-dimensional multi-group radiation hydrodynamics code {\tt STELLA} to calculate light curves because it solves radiation hydrodynamics with non-equilibrium prescription between gas and radiation \citep{Blinnikov1993-vi, Blinnikov1998-ye, Blinnikov2000-xq}, which is required in the low-density regions, \ie the SAGB envelope and CSM. 
{\tt STELLA} solves time-dependent equations about the angular moments of averaged intensity over up to 100 fixed-frequency bins, providing an spectral energy distribution (SED)  at each timestep.
We adopted the standard 100 frequency bins ranging over $1-50,000$ \AA.
Bolometric light curves are produced, integrating the SED at each time step.
Light curves with any filter could be calculated by convolving the SED with the corresponding response function.
In this paper, we provide light curves for the $u$-, $g$-, and $r$-bands of Sloan Digital Sky Survey  \citep{Fukugita1996-ss}, and $B$- and $V$-bands of Johnson-Cousins \citep{Bessell2005-aw}.

\section{Results \label{sec:results}}
First, we present the representative light curves of ECSNe and low-mass FeCCSNe adopting the typical CSM profiles and explosion energies, 
in Section~\ref{subsec:typicalLC}. 
We then present the light curves resulting from wide ranges of physical parameters, varying CSM profile, and explosion energy
in Section~\ref{subsec:allLC}.

\subsection{Representative Light Curves \label{subsec:typicalLC}}
\begin{figure*}[ht!]
\centering
\includegraphics[width=160truemm]{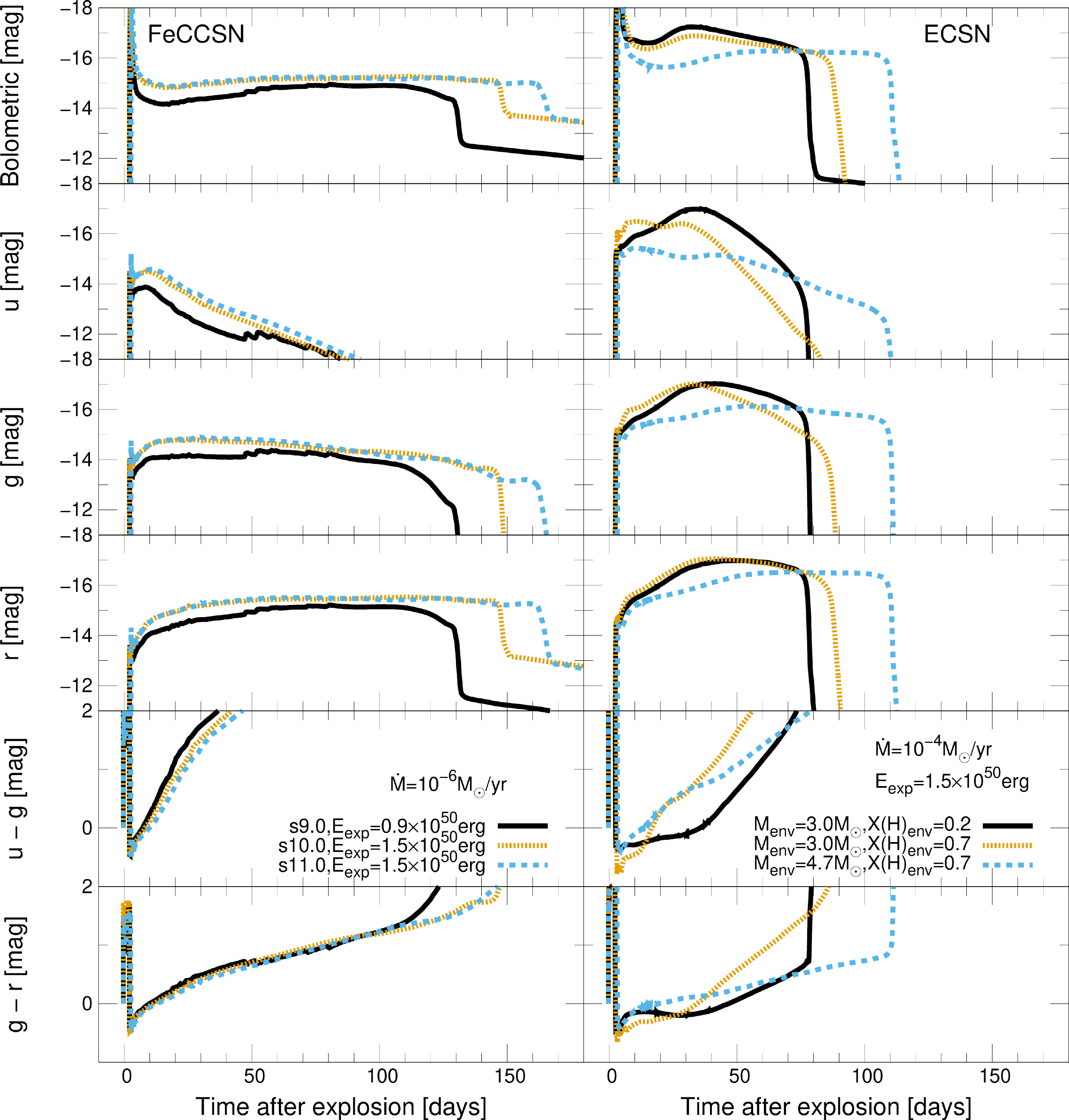}
\caption{Bolometric, $u$-band, $g$-band, and $r$-band light curves and $u-g$ and $g-r$ color evolution of low-mass FeCCSNe and ECSNe of representative progenitors, explosion energy, and CSM.
\label{fig:typiaclLC}}
\end{figure*}

Figure~\ref{fig:typiaclLC} shows the light curves of ECSNe and low-mass FeCCSNe from representative progenitors with typical explosion energies and CSM structure.
The left-hand panels show the bolometric, $u$-band, $g$-band, and $r$-band light curves and the $u-g$ and $g-r$ evolution of low-mass FeCCSNe, and the right-hand panels show those of ECSNe.

The representative FeCCSNe of $\Mms=10.0$ and $11.0~\Msun$ show long ($\sim 150-160$ days) and faint ($\sim -15$ mag in bolometric) plateaus, and the lower-mass FeCCSN of $\Mms=9.0~\Msun$ shows a shorter but still long ($\sim 130$ days) and fainter ($\sim -14$ to $-15$~mag in bolometric) plateau.

The ECSNe have shorter and brighter plateaus than the low-mass FeCCSNe. 
The duration and brightness of plateaus range $70-110$ days and $-16$ to $-17$ mag in bolometric, respectively. The plateaus of ECSNe are longer and slightly fainter if the $\Menv$ is larger or $\XH$ is higher. 
The short and bright plateaus of ECSNe stem from the low-density and extended envelopes of SAGB stars according to the scaling derived in \citet{Popov1993-cu, Eastman1994-hs}.

Furthermore, the ECSNe show bluer plateaus than the FeCCSNe. 
FeCCSNe show a rapid color evolution to red ($\gtrsim 0.06$ mag/day in $u-g$) from just after the shock breakout until approximately $30$ days after explosion. In contrast, ECSNe show a slow color evolution to red ($\lesssim 0.04$ mag/day in $u-g$) for the $\Menv=3.0~\Msun$ and $\XH=0.7$ model and the $\Menv=4.7~\Msun$ and $\XH=0.7$ model, and almost a constant color for the $\Menv=3.0~\Msun$ and $\XH=0.2$ model from just after the shockbreakout until approximately $30$ days after explosion.
This is consistent with the results of \citet{Kozyreva2021-ci}.

\subsection{Light Curves with Wide Physical Quantity Ranges
\label{subsec:allLC}}
In this section, we provide and compare the light curves of ECSNe and low-mass FeCCSNe
with wide ranges of physical quantities
for three cases, depending on the presence and duration of the CSM interaction: models without strong CSM interaction in Section~\ref{subsubsec:noCSMdeg}, models with CSM interaction shorter than the plateau duration in Section~\ref{subsubsec:denseCSMr15deg}, and models with CSM interaction longer than the plateau duration in Section~\ref{subsubsec:denseCSMr16deg}.

\subsubsection{Models Without Strong CSM Interaction
\label{subsubsec:noCSMdeg}}
\begin{figure*}[ht!]
\centering
\includegraphics[width=160truemm]{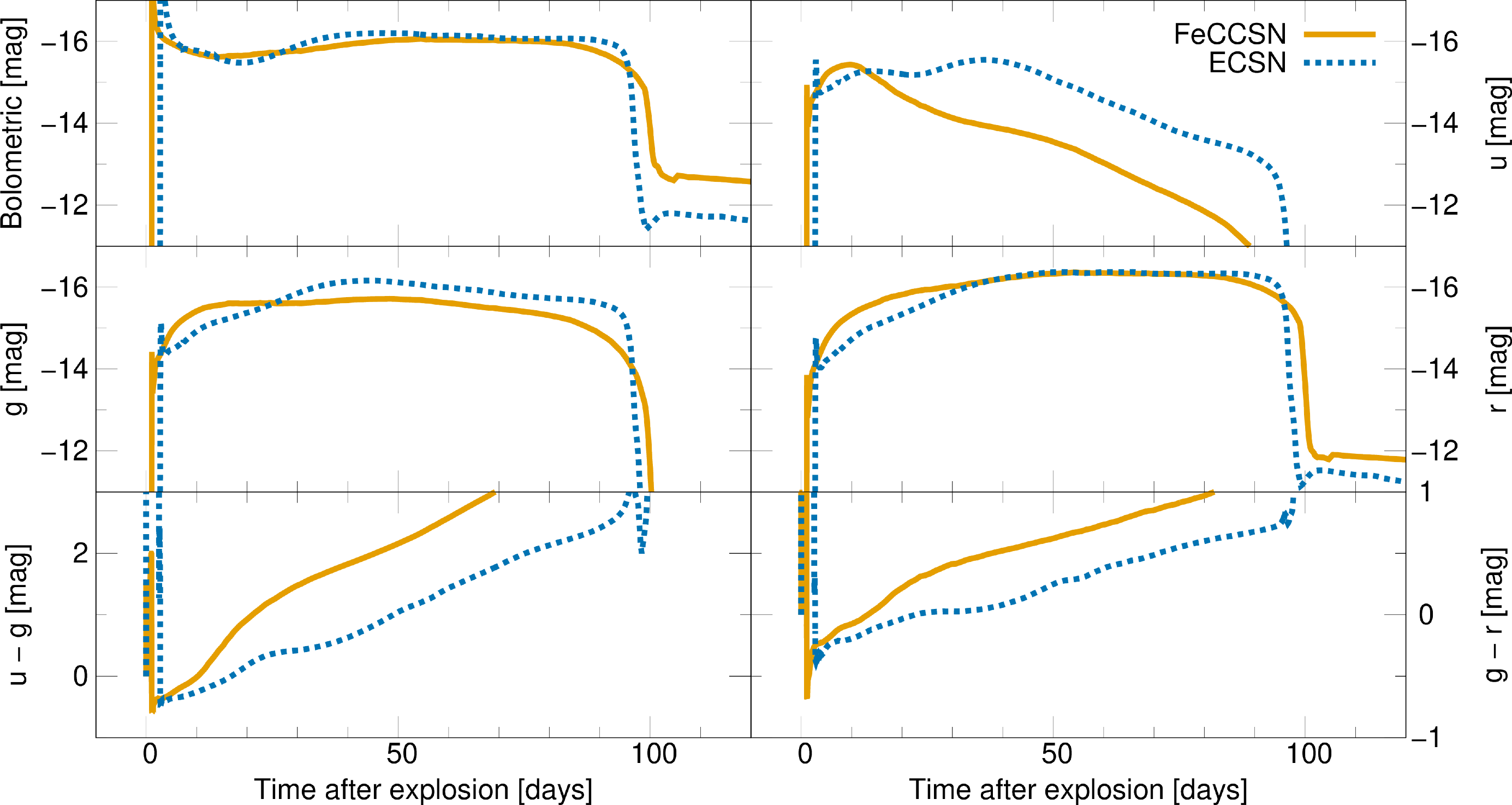}
\caption{
Bolometric, $u$-band, $g$-band, and $r$-band light curves and the $u-g$ and $g-r$ evolution of an ECSN and FeCCSN showing no strong CSM interaction. 
The ECSN is from an SAGB star of \(\Menv=3.0~\Msun\) and \(\XH=0.7\) exploding with \(\Eexp=1.0\times10^{50}\) erg, and the FeCCSN is from an RSG of $\Mms=9.0~\Msun$ exploding with \(\Eexp=3.6\times10^{50}\) erg. 
In both models, no CSM exists around the progenitors. 
\label{fig:LCdeg}}
\end{figure*}

Figure~\ref{fig:LCdeg} shows bolometric, $u$-band, $g$-band, and $r$-band light curves, and the evolution of $u-g$ and $g-r$ of an ECSN and an FeCCSN showing no strong CSM interaction. 
Here, we chose an ECSN from an SAGB star of \(\Menv=3.0~\Msun\) and \(\XH=0.7\) exploding with \(\Eexp=1.0\times10^{50}\) erg, and an FeCCSN from an RSG of $\Mms=9.0~\Msun$ exploding with \(\Eexp=3.6\times10^{50}\) erg. 
In both models, no CSM exists around the progenitors. 

They exhibit similar bolometric light curves. 
This implies that it is difficult to distinguish ECSNe from low-mass FeCCSNe solely with bolometric light curves. This could explain why ECSNe have not been identified in previous transient surveys.
The relations of the explosion energy $\Eexp$ and envelope mass $\Menv$ between ECSN and FeCCSN giving similar bolometric light curves are discussed in Appendix \ref{appsec:boldeg}.

On the other hand, the ECSN and FeCCSN show different multicolor light curves.
The ECSN is brighter than the FeCCSN in the blue bands ($u$- and $g$-band) with similar light curves in the $r$-band. 
The difference is more prominent in the $u$-band than in the $g$-band.
As a result, the ECSN is bluer in the $u-g$ and $g-r$ than the FeCCSN, similar to the representative light curves 
(Section~\ref{subsec:typicalLC}).

\subsubsection{Models with Short and Strong CSM Interaction 
\label{subsubsec:denseCSMr15deg}}
\begin{figure*}[ht!]
\centering
\includegraphics[width=160truemm]{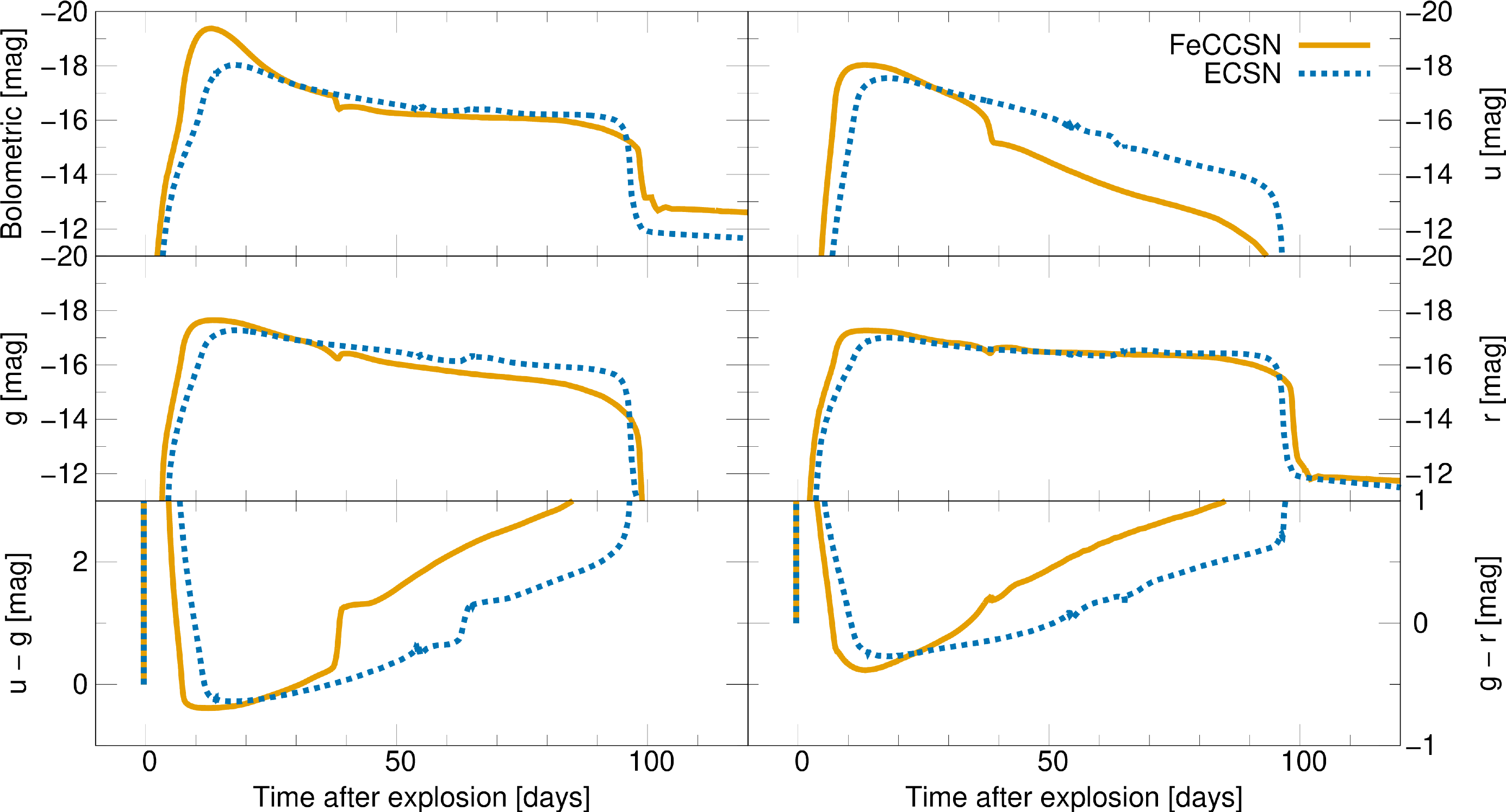}
\caption{
Bolometric, $u$-, $g$-, and $r$-band light curves and the $u-g$ and $g-r$ evolution of an ECSN and FeCCSN showing strong CSM interaction that ends before the plateau does. 
The ECSN is from an SAGB star of \(\Menv=3.0~\Msun\) and \(\XH=0.7\) exploding with \(\Eexp=1.0\times10^{50}\) erg with a CSM of \(\Mdot=10^{-2}~\Myr\), \(\Rcsm=10^{15}\) cm, \(\vi=5\) km/s, \(\vf=10\) km/s, and \(\beta=3\). The FeCCSN is from an RSG of $\Mms=9.0~\Msun$ exploding with \(\Eexp=3.6\times10^{50}\) erg with a CSM of the same parameters as the ECSN. 
\label{fig:LCdenseCSM}}
\end{figure*}

Figure~\ref{fig:LCdenseCSM} shows bolometric, $u$-band, $g$-band, and $r$-band light curves, and the $u-g$ and $g-r$ evolution of an ECSN and FeCCSN with short and strong CSM interaction. 
We adopted the same progenitors and explosion energies as the models in Section~\ref{subsubsec:noCSMdeg} but attached dense and marginally extended CSM.
We adopted the ECSN from an SAGB star of \(\Menv=3.0~\Msun\) and \(\XH=0.7\) exploding with \(\Eexp=1.0\times10^{50}\) erg with a CSM of \(\Mdot=10^{-2}\) $\Myr$, \(\Rcsm=10^{15}\) cm, \(\vi=5\) km/s, \(\vf=10\) km/s, and \(\beta=3\), and the FeCCSN from an RSG of $\Mms=9.0~\Msun$ exploding with \(\Eexp=3.6\times10^{50}\) erg with a CSM of the same parameters as the ECSN. 

The ECSN shows a fainter bolometric peak than the FeCCSN, where the interaction between the SN ejecta and CSM enhances the luminosity. Their light curves are similar in the later epochs after the CSM interaction ends (approximately 60 days for the ECSN and 40 days for the FeCCSN after the explosion).
The similar late plateau results from the lower explosion energy of the ECSN compared to the FeCCSN, as in Section~\ref{subsubsec:noCSMdeg} (Appendix~\ref{appsec:boldeg}). 
The lower explosion energy results in a lower SN ejecta velocity, weaker CSM interaction, and thus a fainter peak.
The ECSN exhibits a bluer color than the FeCCSN after the CSM interaction until the plateau ends.

\subsubsection{Models with Long and Strong CSM Interaction
\label{subsubsec:denseCSMr16deg}}
\begin{figure*}[ht!]
\centering
\includegraphics[width=160truemm]{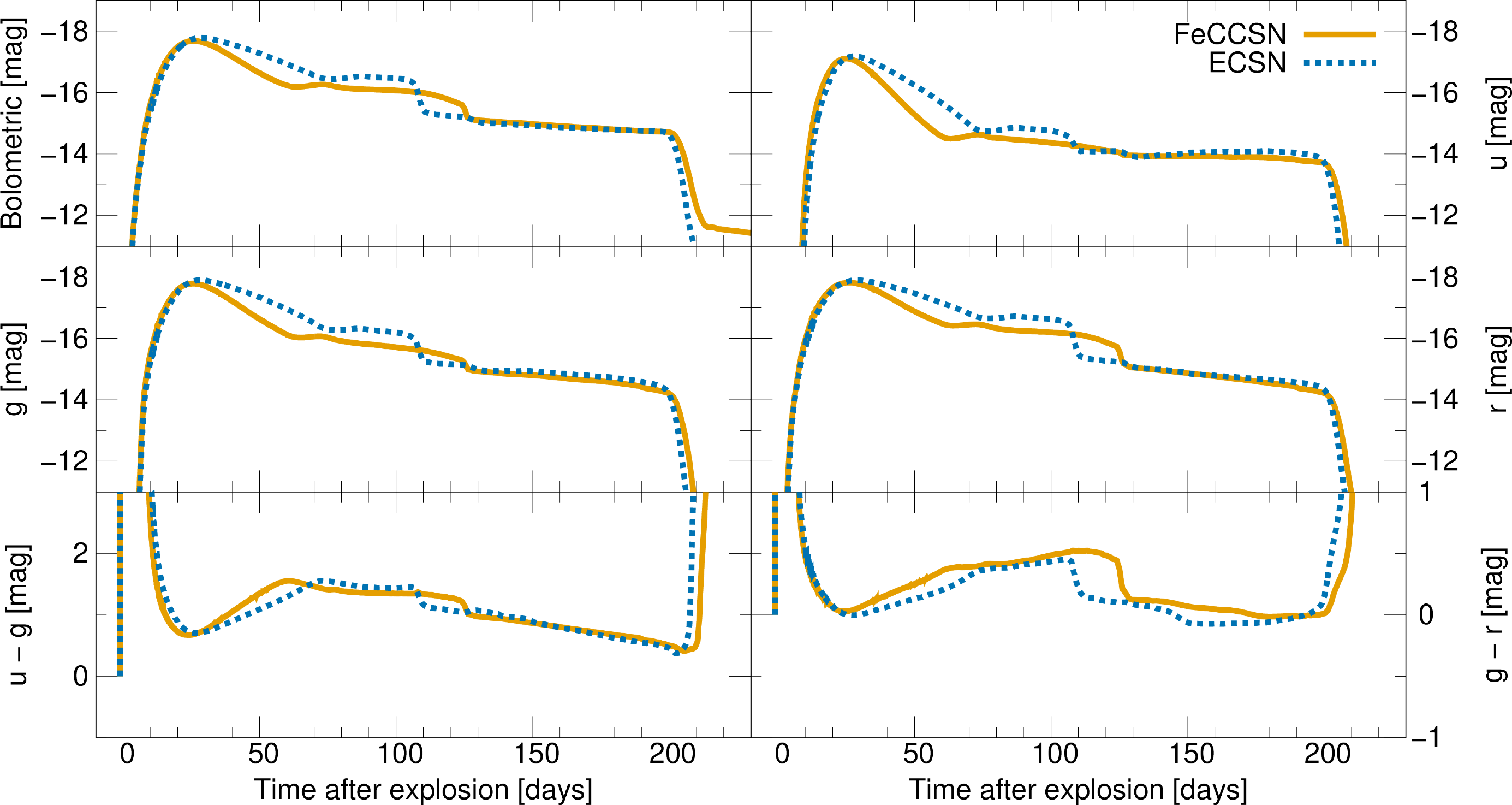}
\caption{
Bolometric, $u$-band, $g$-band, and $r$-band light curves and the $u-g$ and $g-r$ evolution of an ECSN and FeCCSN showing strong CSM interaction lasting after the plateau ends. 
The ECSN is from an SAGB star of \(\Menv=3.0~\Msun\) and \(\XH=0.7\) exploding with \(\Eexp=1.0\times10^{50}\) erg with a CSM of \(\Mdot=10^{-2}~\Myr\), \(\Rcsm=3\times10^{15}\) cm, \(\vi=5\) km/s, \(\vf=10\) km/s, and \(\beta=3\), and the FeCCSN is from an RSG of $\Mms=9.0~\Msun$ exploding with \(\Eexp=1.4\times10^{50}\) erg with a CSM of the same parameters as the ECSN.
\label{fig:LCdenseCSMr3d15}}
\end{figure*}

Figure~\ref{fig:LCdenseCSMr3d15} shows bolometric, $u$-band, $g$-band, and $r$-band light curves, and the $u-g$ and $g-r$ evolution of an ECSN and an FeCCSN showing strong and long-lasting CSM interaction. 
We chose an ECSN from an SAGB star of \(\Menv=3.0~\Msun\) and \(\XH=0.7\) exploding with \(\Eexp=1.0\times10^{50}\) erg and an FeCCSN from an RSG of $\Mms=9.0~\Msun$ exploding with \(\Eexp=1.4\times10^{50}\) erg. 
We adopted a CSM with the same parameters in both models: \(\Mdot=10^{-2}\) $\Myr$, \(\Rcsm=3\times10^{15}\) cm, \(\vi=5\) km/s, \(\vf=10\) km/s, and \(\beta=3 \).

The ECSN and FeCCSN have similar bolometric and multicolor light curves. The reason for this is discussed in Appendix~\ref{appsec:deg_typeiii}. In contrast to the models in Section~\ref{subsubsec:noCSMdeg} and \ref{subsubsec:denseCSMr15deg}, no color difference between the ECSN and FeCCSN are found in these models.

\section{Discussion \label{sec:discussion}}
\subsection{The Blue Plateau of ECSNe \label{subsec:bluePT}}
In this section, we discuss the origin of the bluer color of ECSNe than FeCCSNe found in Section~\ref{subsubsec:noCSMdeg} and \ref{subsubsec:denseCSMr15deg}.
Here we chose the ECSN from an SAGB star of $\Menv = 3.0~\Msun$ and $\XH = 0.7$ exploding with $\Eexp=1.0\times10^{50}$ erg, and the FeCCSN from an RSG of $\Mms=9.0~\Msun$ exploding with $\Eexp=3.6\times10^{50}$ erg shown in Figure~\ref{fig:LCdeg}. In both models, no CSM exists around the progenitors so that we could investigate the intrinsic color.

\begin{figure}[ht!]
\centering
\includegraphics[width=80truemm]{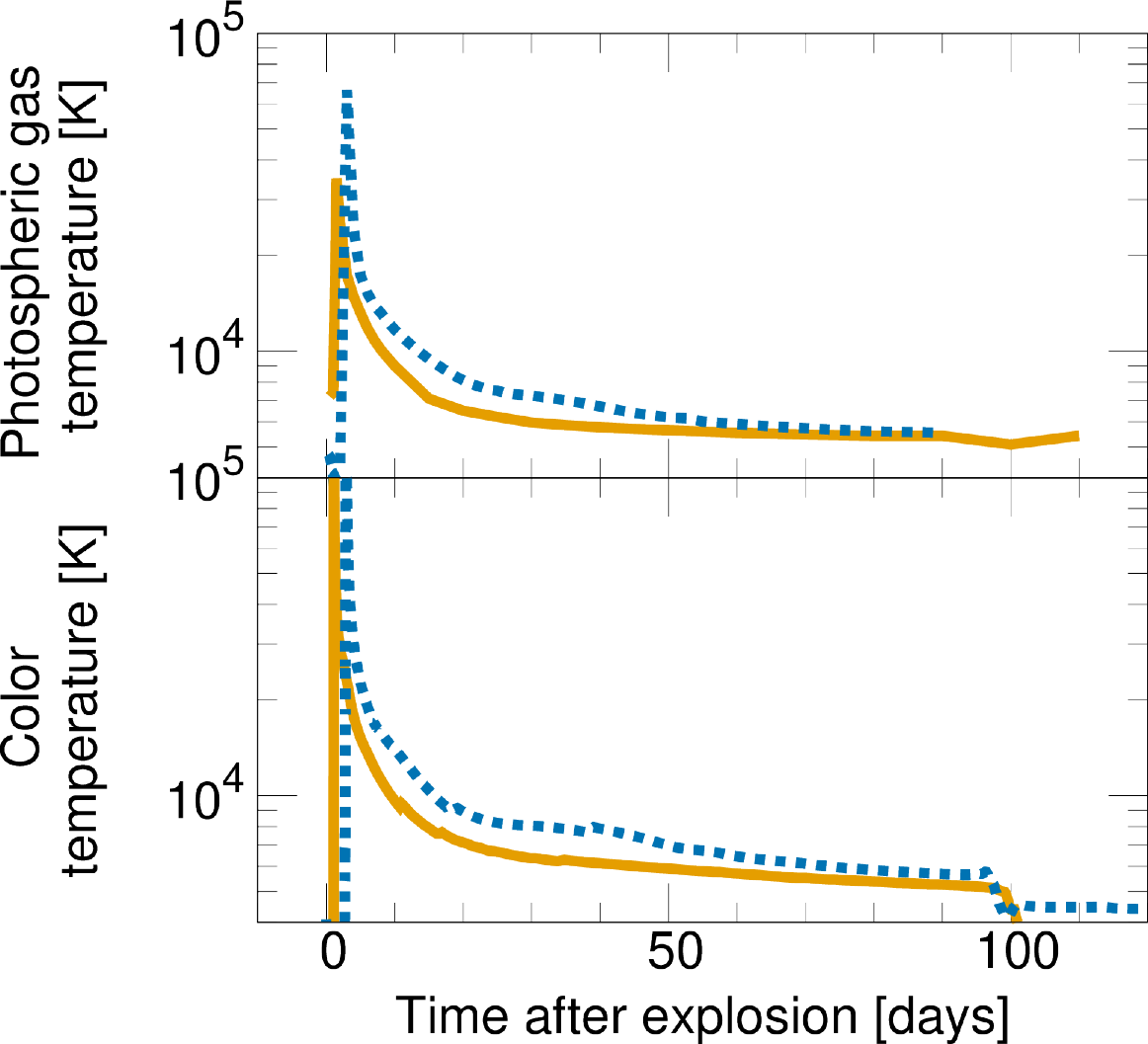}
\caption{
Evolution of the photospheric temperature $\Tph$ and color temperature $\Tcol$ of the ECSN and FeCCSN, for which light curves are shown in Figure~\ref{fig:LCdeg}. 
The ECSN is from an SAGB star of \(\Menv=3.0~\Msun\) and \(\XH=0.7\) exploding with $\Eexp=1.0 \times 10^{50}$ erg, and the FeCCSN is from an RSG of $\Mms=9.0~\Msun$ exploding with \(\Eexp=3.6\times10^{50}\) erg. 
In both models, no CSM exists around the progenitors. 
\label{fig:Tevl}}
\end{figure}

Figure~\ref{fig:Tevl} shows the evolution of the photospheric temperature $\Tph$ and color temperature $\Tcol$ of the ECSN and FeCCSN.
While $\Tph$ and $\Tcol$ decline with time in both the ECSN and FeCCSN, those of the ECSN are higher than those of the FeCCSN. 
This indicates that the higher $\Tph$ of the ECSN causes the bluer color. 

\begin{figure*}[ht!]
\centering
\includegraphics[width=140truemm]{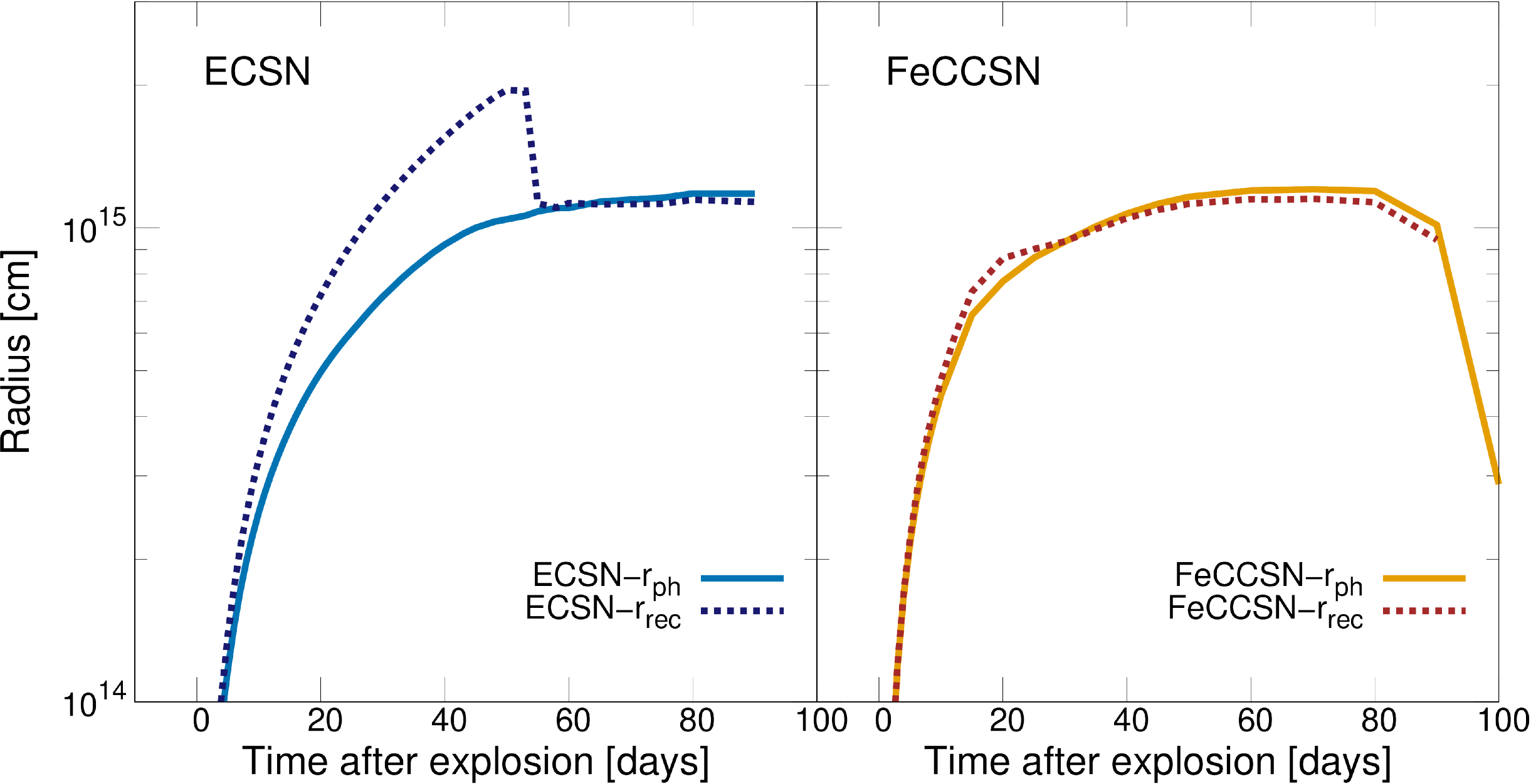}
\caption{Time evolution of the photospheric radius $\rph$ and the recombination front $\rrec$ of the ECSN and FeCCSN in Figure~\ref{fig:LCdeg}.
\label{fig:rph_and_recf}}
\end{figure*}

Figure~\ref{fig:rph_and_recf} shows the time evolution of the photospheric radii $\rph$ and recombination front radii $\rrec$ of the ECSN and FeCCSN.
The recombination front $\rrec$ is defined as the outermost radius where the ionization fraction $x$ is higher than $0.1$.
During the plateau, the photosphere of the ECSN is mostly located deeper than the recombination front ($\rph < \rrec$, until approximately 55 days after the explosion), whereas the radii of the photosphere and the recombination front of the FeCCSN are almost same ($\rph \simeq \rrec$, after approximately 20 days after the explosion). 

\begin{figure*}[ht!]
  \begin{minipage}[b]{0.45\linewidth}
    \centering
    \includegraphics[width=80truemm]{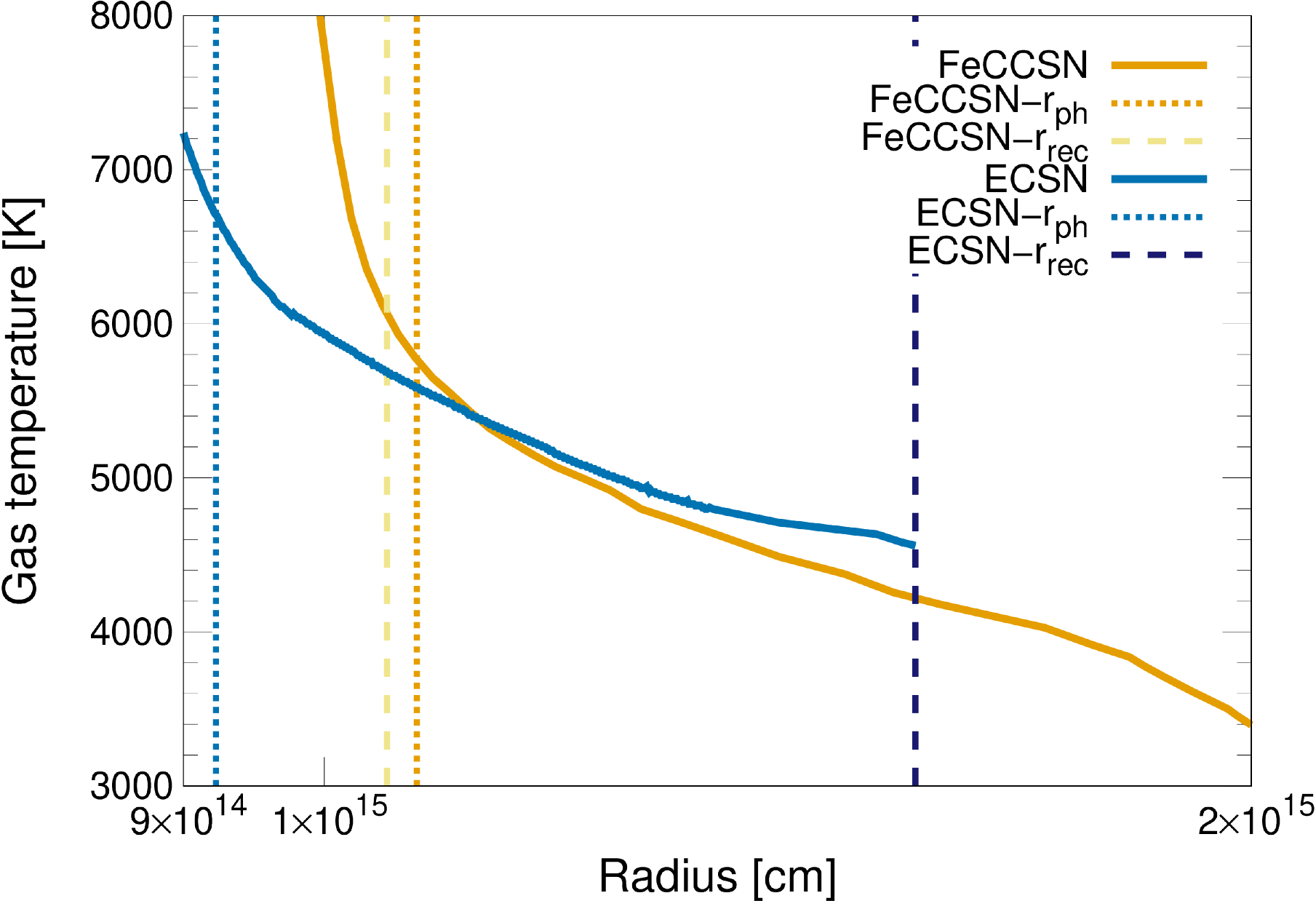}
  \end{minipage}
  \begin{minipage}[b]{0.45\linewidth}
    \centering
    \includegraphics[width=80truemm]{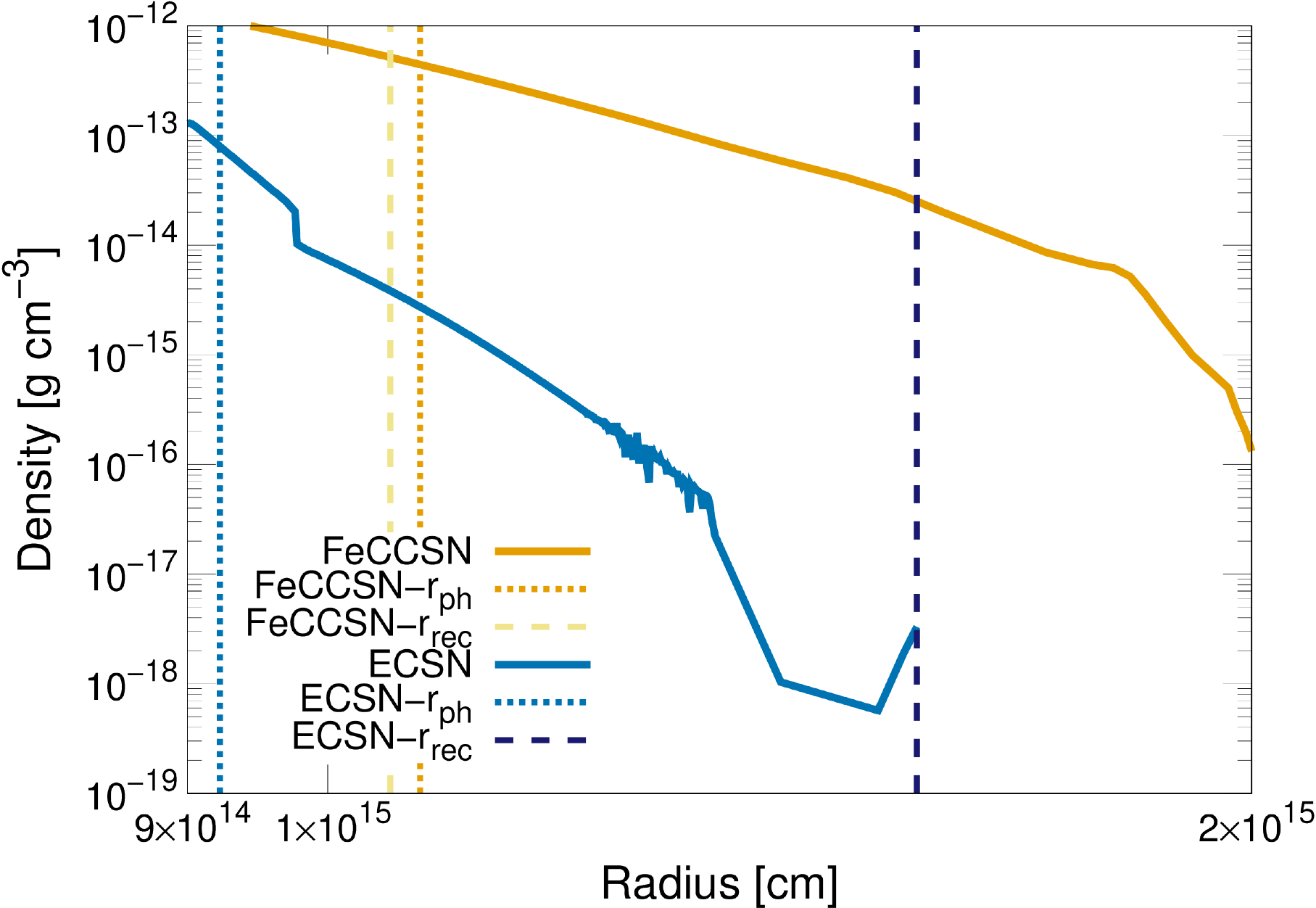}
  \end{minipage}
  \caption{
  Temperature (left panel) and density (right panel) structures of the ejecta of the ECSN and FeCCSN in Figure~\ref{fig:LCdeg} at 40 days after the explosion. The radii of the photosphere \(\rph\) and the recombination front \(\rrec\) are indicated by vertical dotted and dashed lines, respectively.
  \label{fig:r_rho_Tg}}
\end{figure*}

The left panel of Figure~\ref{fig:r_rho_Tg} shows the temperature structures of the ejecta of the ECSN and FeCCSN at 40 days after the explosion. 
The ECSN has a higher $\Tph$ than recombination temperature as a result of the relation $\rph < \rrec$, while the FeCCSN has almost the same $\Tph$ and recombination temperature because of the relation $\rph \simeq \rrec$. 

The right panel of Figure~\ref{fig:r_rho_Tg} shows the density structures of the ejecta of the ECSN and FeCCSN at 40 days after the explosion. 
The ECSN has a largely low density around the recombination front $\rrec$
resulting in $\rph < \rrec$.
Conversely, the FeCCSN has high density around $\rrec$, resulting in $\rph \simeq \rrec$.
Therefore, the blue plateau of the ECSN is caused by the low-density ejecta of the ECSN, which originates from the low-density envelope of the SAGB star.

\citet{Kozyreva2021-ci} also show the blue plateau of ECSNe and claim that the reason is because the recombination of ECSN hardly settles because of the low-density envelope of a SAGB star, making the photosphere recede slowly. 
Indeed, 
we can find that the $\rrec$ of the ECSN recedes more slowly than that of the FeCCSN (Figure~\ref{fig:rph_and_recf}), agreeing that recombination of the ECSN hardly settles. 
However, we find that the $\rph$ and $\rrec$ of the ECSN do not match in fact.
The $\rph$ of the ECSN is located deeper than the $\rrec$, resulting in the bluer color of the ECSN.

\subsection{ECSN Diagnostic Method \label{subsec:identification}}
In this section, we present a new diagnostic method for ECSNe.
First, we define a pair of observational quantities:
\begin{enumerate}
    \item \(t_0\): a time at which the $g$-band absolute magnitude reaches $-13$ mag, as presented in \citet{Moriya2018-ve}. \(t_0\) is set as the origin of time in this section.
    \item \(\tPT\): the time of the mid-point between the end of the plateau and the start of the tail. \(\tPT\) is derived by fitting equation~(1) of \citet{Valenti2016-ao} to the calculated $g$-band light curve.
\end{enumerate}

\begin{figure*}[ht!]
  \begin{minipage}[b]{0.45\linewidth}
    \centering
    \includegraphics[width=80truemm]{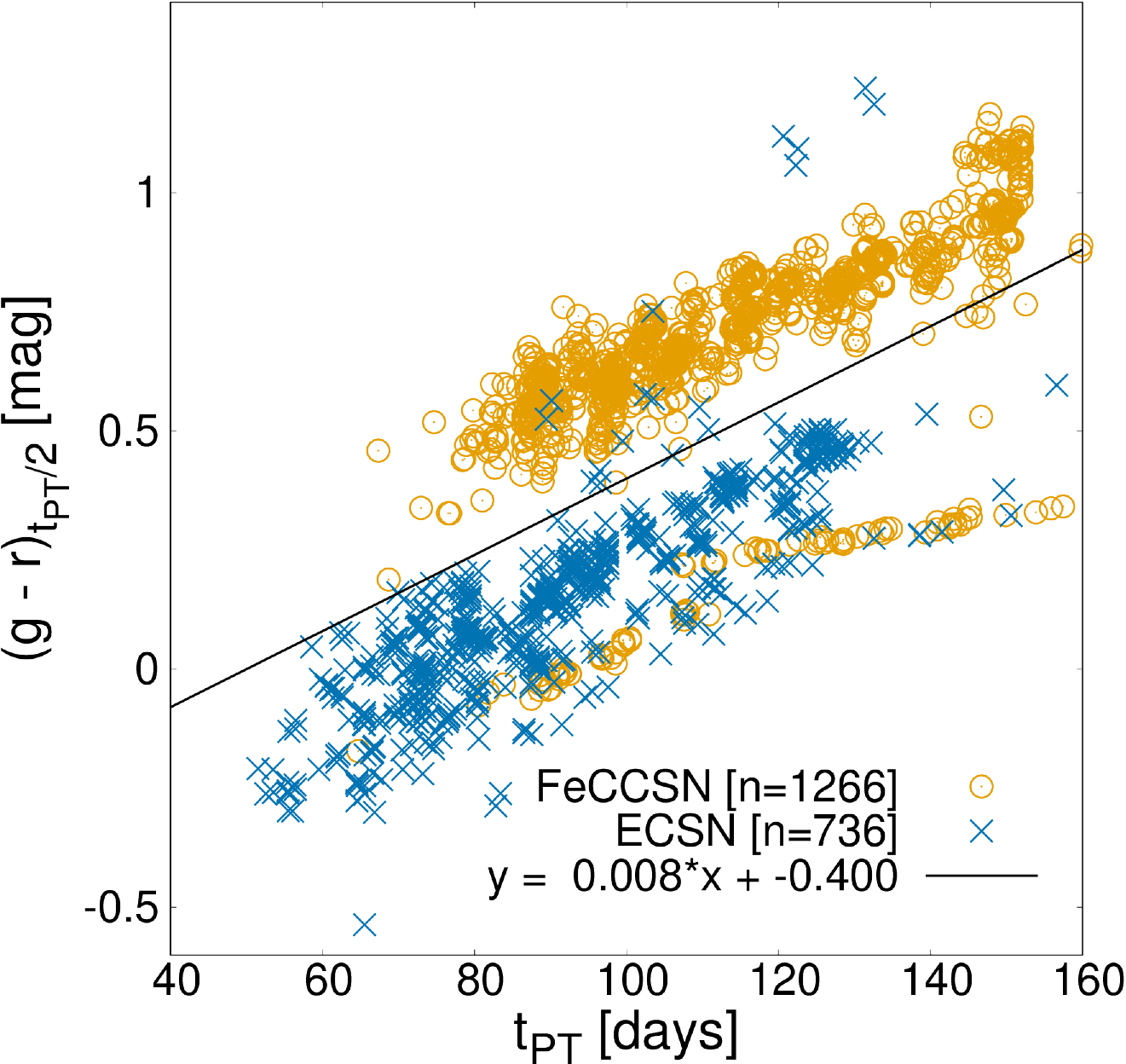}
  \end{minipage}
  \begin{minipage}[b]{0.45\linewidth}
    \centering
    \includegraphics[width=80truemm]{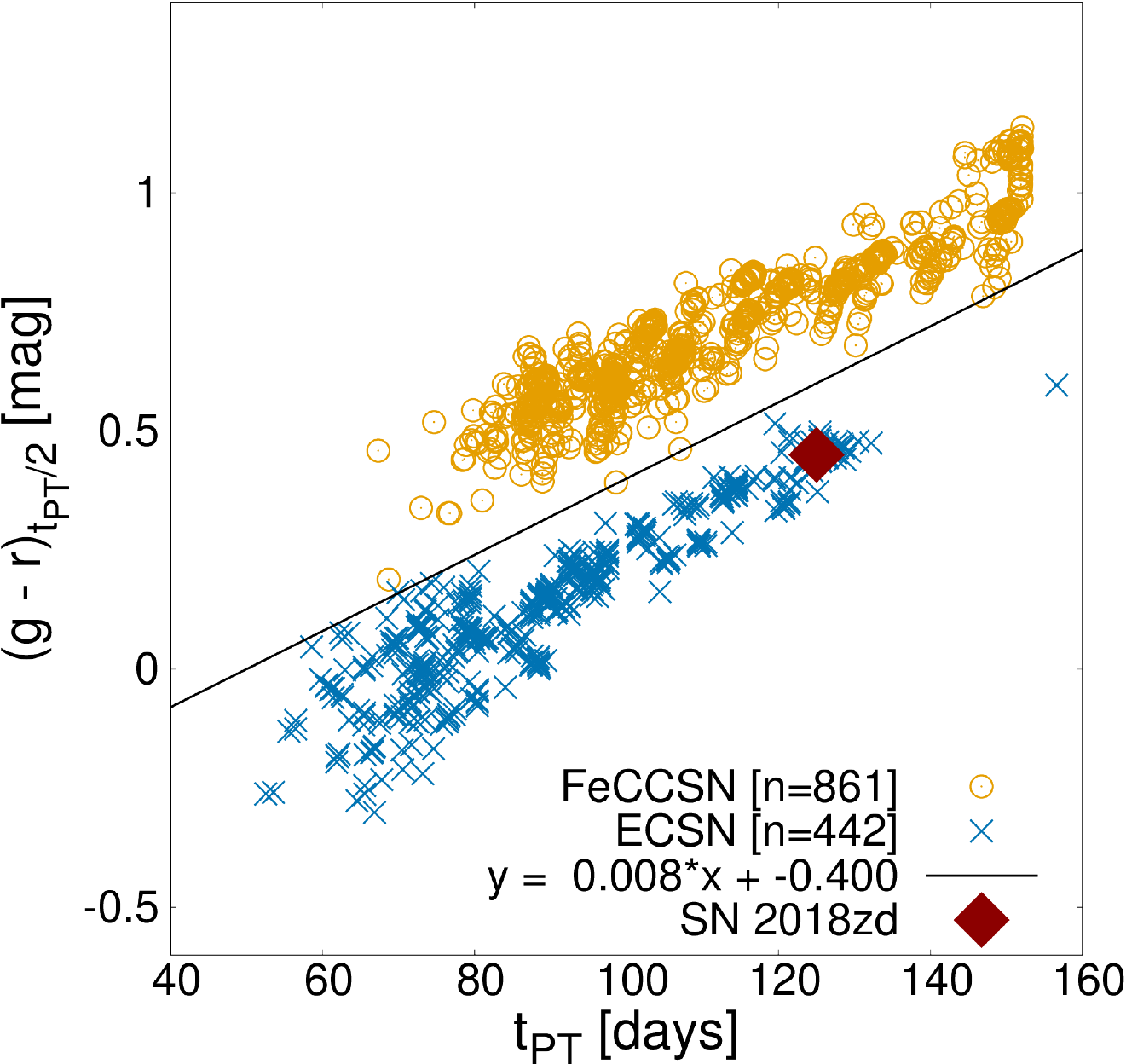}
  \end{minipage}
  \caption{Diagnostic method to select ECSNe. The horizontal axis shows \(\tPT\) and the vertical axis shows the $g-r$ color at \(\tPT\)/2 for all models (left panel) and for models showing no strong CSM interaction at $\tPT/2$ (right panel). SN~2018zd is overplotted in the right panel with a dark-red diamond.
  \label{fig:tpt_grhalf}}
\end{figure*}

Figure~\ref{fig:tpt_grhalf} shows the relation between $\tPT$ and the $g-r$ color at $t=\tPT/2$ [$(g-r)_{\tPT/2}$].
The points indicate ECSN models (blue crosses) and FeCCSN models (orange circles). 
While the left panel shows all the models (736 ECSNe and 1266 FeCCSNe) calculated in this work, the right panel shows the models not showing strong CSM interaction at \(\tPT/2\) (442 ECSNe and 861 FeCCSNe), that is, the models without strong CSM interaction (Section~\ref{subsubsec:noCSMdeg}) and part of the models with short and strong CSM interaction (Section~\ref{subsubsec:denseCSMr15deg}). 
In the right panel, the ECSNe are clearly separated from the FeCCSNe and located at 
\begin{equation}
(g-r)_{\tPT/2} < 0.008 \times \tPT - 0.4,
\label{eq:g-r(tPT/2)threshold}
\end{equation}
reflecting their bluer plateau.
This is a novel diagnostic method because it is independent of the distance.

We apply our diagnostic method to SN~2018zd, which is proposed as an ECSN \citep{Hiramatsu2021-er} and a low-mass FeCCSN \citep{Callis2021-zu}. The disagreement is caused by uncertainty in the distance to SN~2018zd.
The diagnostic method can be applied because the spectral evolution of SN~2018zd (Extended Data Figure~4 of \citealt{Hiramatsu2021-er}) indicates that CSM interaction ends earlier than \(\tPT\)/2. 
We estimate \(\tPT=125\)~days and $(g-r)_{\tPT/2}=0.45$. SN~2018zd is shown with a dark-red diamond in the right panel of Figure~\ref{fig:tpt_grhalf}.
SN~2018zd clearly satisfies the criterion of equation~(\ref{eq:g-r(tPT/2)threshold}). This indicates that SN~2018zd is an ECSN, according to the distance-independent diagnostic method.

We present alternative indicators, in which $B-V$ is adopted instead of $g-r$ and several epochs (\eg $t=40$ days) are adopted instead of $\tPT/2$, in Appendix~\ref{appsec:other_diagnostics}.
We propose a procedure of observations to find ECSNe in Appendix~\ref{appsec:obs_procedure}.

\section{Conclusion \label{sec:conclusion}}
We have presented multicolor light curves of ECSNe and low-mass FeCCSNe. 
We first adopted the physical quantities estimated in previous studies with theoretically or observationally plausible methods 
to reveal representative light curves, 
and then adopted wide physical quantity ranges
to investigate the robustness of their intrinsic differences. 
We adopted SAGB progenitor models with the envelope masses of $\Menv=2.0-4.7~\Msun$ from \citet{Tominaga2013} for ECSNe and RSG progenitor models with the zero-age main-sequence masses of $\Mms=9-12~\Msun$ (s9.0--s12.0) from \citet{Sukhbold2016} for FeCCSNe. We also considered various CSM structures. We used the multi-group radiation hydrodynamics code, {\tt STELLA} \citep{Blinnikov1993-vi, Blinnikov1998-ye, Blinnikov2000-xq} to calculate light curves. 

The representative light curves of ECSNe show brighter and shorter plateaus than those of low-mass FeCCSNe.
This stems from the low-density and extended envelopes of SAGB stars according to the scaling derived in \citet{Popov1993-cu, Eastman1994-hs}.

The calculations with wide physical quantity ranges
revealed that an ECSN and a low-mass FeCCSN can display similar bolometric light curves. 
This may explain why we could not identify ECSNe in previous transient surveys. 
ECSNe show bluer plateaus than FeCCSNe 
if no strong CSM interaction is taking place at the epoch.

The bluer plateaus of ECSNe originate from the low-density envelopes of SAGB stars. 
The low-density envelope results that the photosphere of an ECSN is mostly located deeper than the recombination front during the plateau, whereas the radii of the photosphere and the recombination front are almost same in an FeCCSN.
Therefore, the photospheric temperature of an ECSN is higher than the recombination temperature, whereas the photospheric temperature of an FeCCSN is almost same as the recombination temperature. This causes the bluer plateaus of ECSNe than those of FeCCSNe.

We proposed a new diagnostic method to select ECSNe. 
In the method, the transition time from plateau to tail phase (\(\tPT\)) and the $g-r$ at $\tPT/2$ [$(g-r)_{\tPT/2}$] are adopted. 
ECSNe show bluer color than low-mass FeCCSNe at \(\tPT\)/2 
in the absence of a strong CSM interaction.
The novel aspect of this method is that we can distinguish ECSNe from low-mass FeCCSNe independently of distance.

We apply the method to SN~2018zd, which has been proposed as a promising ECSN candidate \citep{Hiramatsu2021-er} and as a low-mass FeCCSN \citep{Callis2021-zu}. 
The disagreement between the propositions is primarily caused by the different estimation of its distance, which results in different ejected \Nifs~mass estimations. Our new method indicates that SN~2018zd is an ECSN.

We propose a procedure of observations to find ECSNe: 1) explore SNe~II showing blue-colored multicolor light curves, 2) check if the strong CSM interaction is taking place using spectroscopic observations, and 3) diagnose if it is an ECSN or not by applying the new method.

\begin{acknowledgments}
We thank Masaomi Tanaka, Keiichi Maeda, and Tomoya Takiwaki for helpful discussions.
Numerical computations were in part carried out on PC cluster at Center for Computational Astrophysics, National Astronomical Observatory of Japan. 
SIB and MSP thank the Russian Foundation for Basic Research (RFBR) and the Deutsche Forschungsgemeinschaft (DFG), Joint RFBR-DFG project number 21-52-12032.
\end{acknowledgments}

\bibliography{main}{}
\bibliographystyle{aasjournal}

\appendix
\section{Why ECSNe and low-mass F\lowercase{e}CCSNe without strong CSM interaction or with short and strong CSM interaction exhibit similar bolometric light curves
\label{appsec:boldeg}}
ECSNe and low-mass FeCCSNe can have similar bolometric light curves if no strong CSM interaction takes place (Section~\ref{subsubsec:noCSMdeg}) or if the strong CSM interaction ends during the plateau (Section~\ref{subsubsec:denseCSMr15deg}).
According to equations (9) and (10) of \citet{Eastman1994-hs}, we derive relations among the explosion energy $\Eexp$, envelope mass $\Menv$, and pre-supernova radius of such an ECSN and low-mass FeCCSN. 
Let us assume the plateau duration \(\taupl\) and luminosity \(\Lpl\) of an ECSN and a low-mass FeCCSN are the same, \ie
\begin{equation}
\tau_{\rm{pl, Fe}} = \tau_{\rm{pl, EC}}
\end{equation}
and 
\begin{equation}
L_{\rm{pl, Fe}}=L_{\rm{pl, EC}},
\end{equation}
where EC and Fe denote ECSN and FeCCSN respectively. 
Using these equations, the relations among the explosion energy, envelope mass, and pre-supernova radius are derived as
\begin{equation}
\label{eq:Eratio_deg}
\frac{E_{\rm{exp, EC}}}{E_{\rm{exp, Fe}}} \sim \left(\frac{R_{\rm{RSG}}}{R_{\rm{SAGB}}}\right)^\frac{5}{4}
\end{equation}
and
\begin{equation}
\label{eq:Mratio_deg}
\frac{M_{\rm{env, EC}}}{M_{\rm{env, Fe}}} \sim \left(\frac{R_{\rm{RSG}}}{R_{\rm{SAGB}}}\right)^\frac{3}{4}.
\end{equation}
The progenitor radii of the ECSN and FeCCSN of the models discussed in Section~\ref{subsubsec:noCSMdeg} are \(7.1 \times 10^{13}\) and \(2.9 \times 10^{13}\) cm, respectively. 
Hence, the ratio of $\Eexp$ is $\sim 0.3$ and the ratio of $\Menv$ is $\sim 0.5$, 
which is consistent with the ratios of their parameters ($\Eexp$ ratio $\sim 0.3$ and $\Menv$ ratio $\sim 0.4$).

\section{Why ECSNe and low-mass F\lowercase{e}CCSNe with long and strong CSM interactions exhibit similar bolometric and multicolor light curves
\label{appsec:deg_typeiii}}
The ECSN and FeCCSN discussed in Section~\ref{subsubsec:denseCSMr16deg} exhibit similar bolometric and multicolor light curves.
The observational quantities which characterize the light curves are peak luminosity $\Lp$, peak color $\Tp$, and the duration of CSM interaction $\taucsm$ if they show long-lasting strong CSM interaction that ends after the plateau does.
$\Lp$ is expressed as \citep{Chevalier2011-ds} 
\begin{equation}
\Lp = \Lp(\Eexp, \Mej, \Mdot).
\end{equation}
$\Tp$ is roughly expressed as 
\begin{equation}
\Tp = \Tp(\Lp(\Eexp, \Mej, \Mdot), \Rcsm),
\end{equation}
if we assume that the radiation is approximately blackbody and that the photosphere is located near the outermost region of the CSM.
$\taucsm$ is roughly expressed as 
\begin{equation}
\taucsm = \taucsm(\Eexp, \Mej, \Rcsm) \propto \frac{\Rcsm}{\sqrt{\Eexp/M_{ej}}}
\end{equation}
because CSM interaction lasts during the shock propagation in the CSM. 
There are three  free quantities independent of the pre-supernova structure among the four physical quantities $\Mej$, $\Eexp$, $\Mdot$, and $\Rcsm$. Only $\Mej$ depends on the pre-supernova structure because ECSNe have lower $\Mej$ than low-mass FeCCSNe.
Therefore, the three observational quantities $\Lp$, $\Tp$, and $\taucsm$ can be same in ECSNe and FeCCSNe if a certain relation of $\Eexp$, $\Mdot$, and $\Rcsm$ is satisfied.

\section{Other indicators of ECSNe
\label{appsec:other_diagnostics}}
\begin{figure}[ht!]
\centering
\includegraphics[width=80truemm]{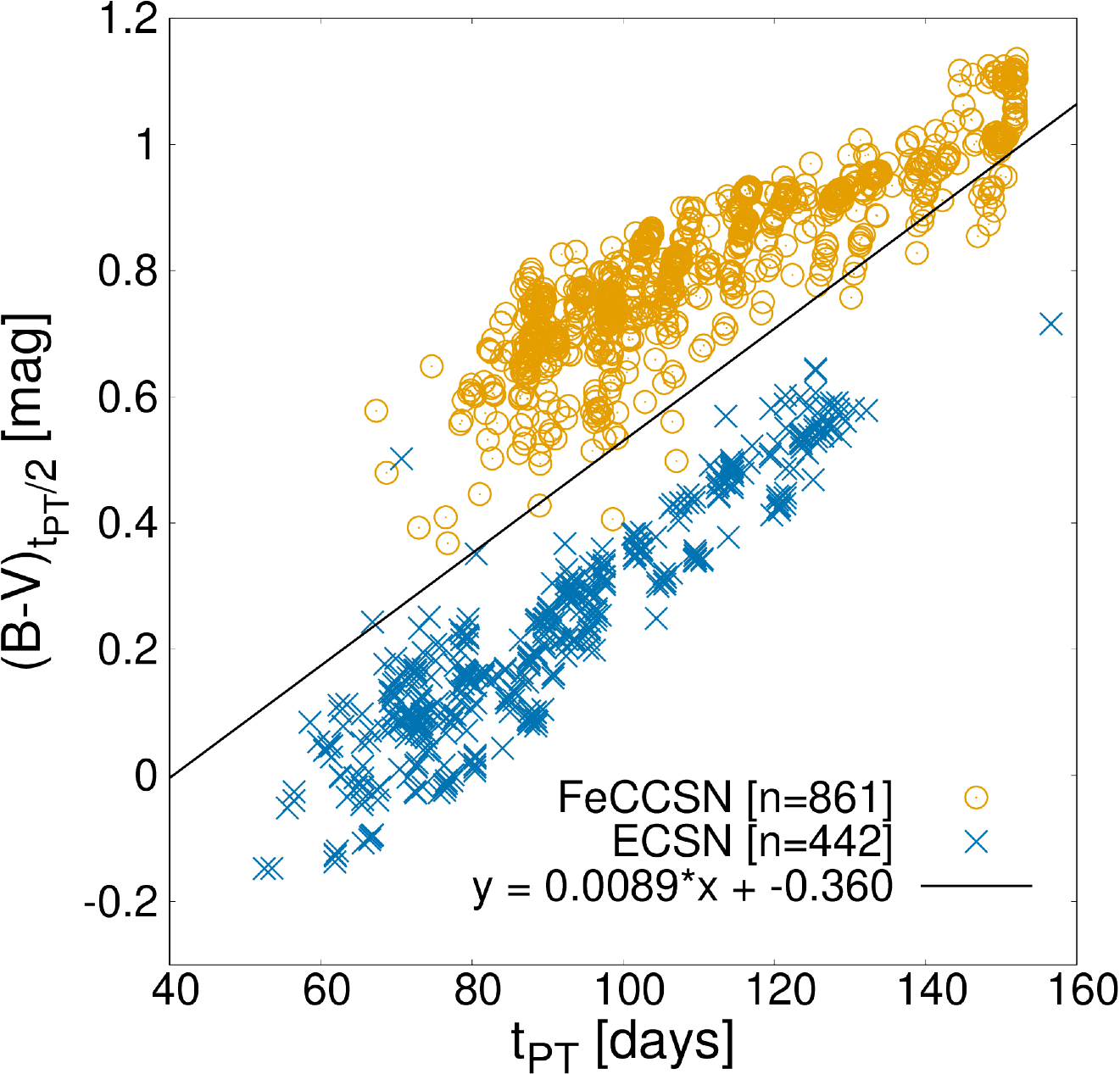}
\caption{
An indicator of ECSNe in which $B-V$ is adopted instead of $g-r$.
\label{fig:tpt_BVhalf}}
\end{figure}

We have presented a new diagnostic method to select ECSNe in Section~\ref{subsec:identification}, in which the transition time from plateau to tail $\tPT$ and the color $g-r$ at $\tPT/2$ are adopted. 
Here, we also present an alternative indicator in which the color index $B-V$ is adopted instead of $g-r$.
Figure~\ref{fig:tpt_BVhalf} shows the indicator. The horizontal axis shows the $\tPT$ and the vertical axis shows $B-V$ at $\tPT/2$ [$(B-V)_{\tPT/2}$]. 
ECSNe can be selected with a criterion, 
\begin{equation}
(B-V)_{\tPT/2} < 0.0089 \times \tPT - 0.36.
\label{eq:B-V(tPT/2)threshold}
\end{equation}

As to be discussed in Appendix~\ref{appsec:obs_procedure}, 
in the procedure of observations, 
we need to know if the CSM interaction ends before $\tPT/2$ from spectroscopic observations although we can only know $\tPT$ after the plateau ends. 
This is crucial, especially when we diagnose far or ubiquitous objects, because frequent spectroscopic observations are difficult.

\begin{figure}[ht!]
\begin{tabular}{ccc}
  \begin{minipage}[b]{0.3\linewidth}
    \centering
    \includegraphics[width=50truemm]{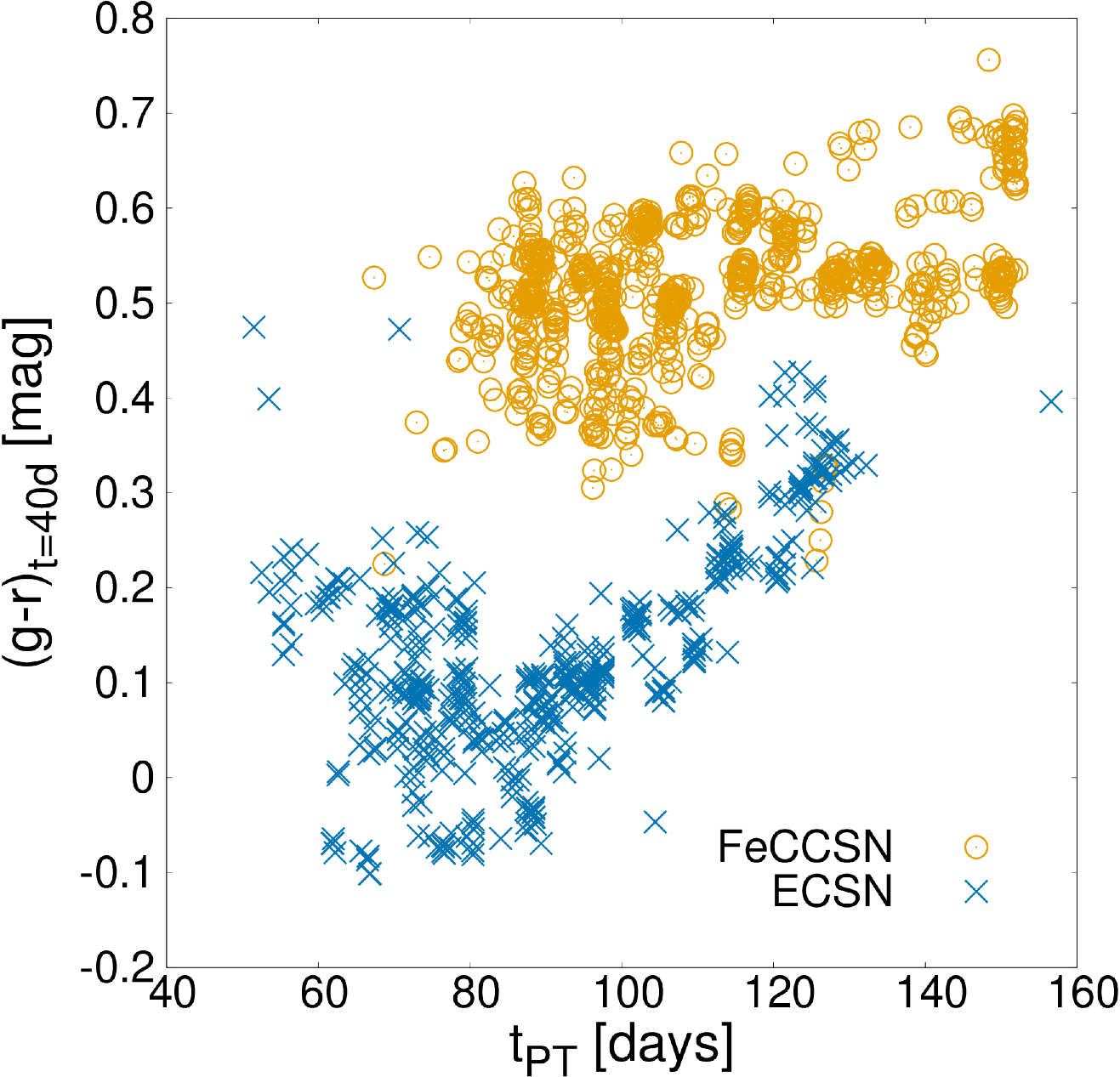}
  \end{minipage} &
  \begin{minipage}[b]{0.3\linewidth}
    \centering
    \includegraphics[width=50truemm]{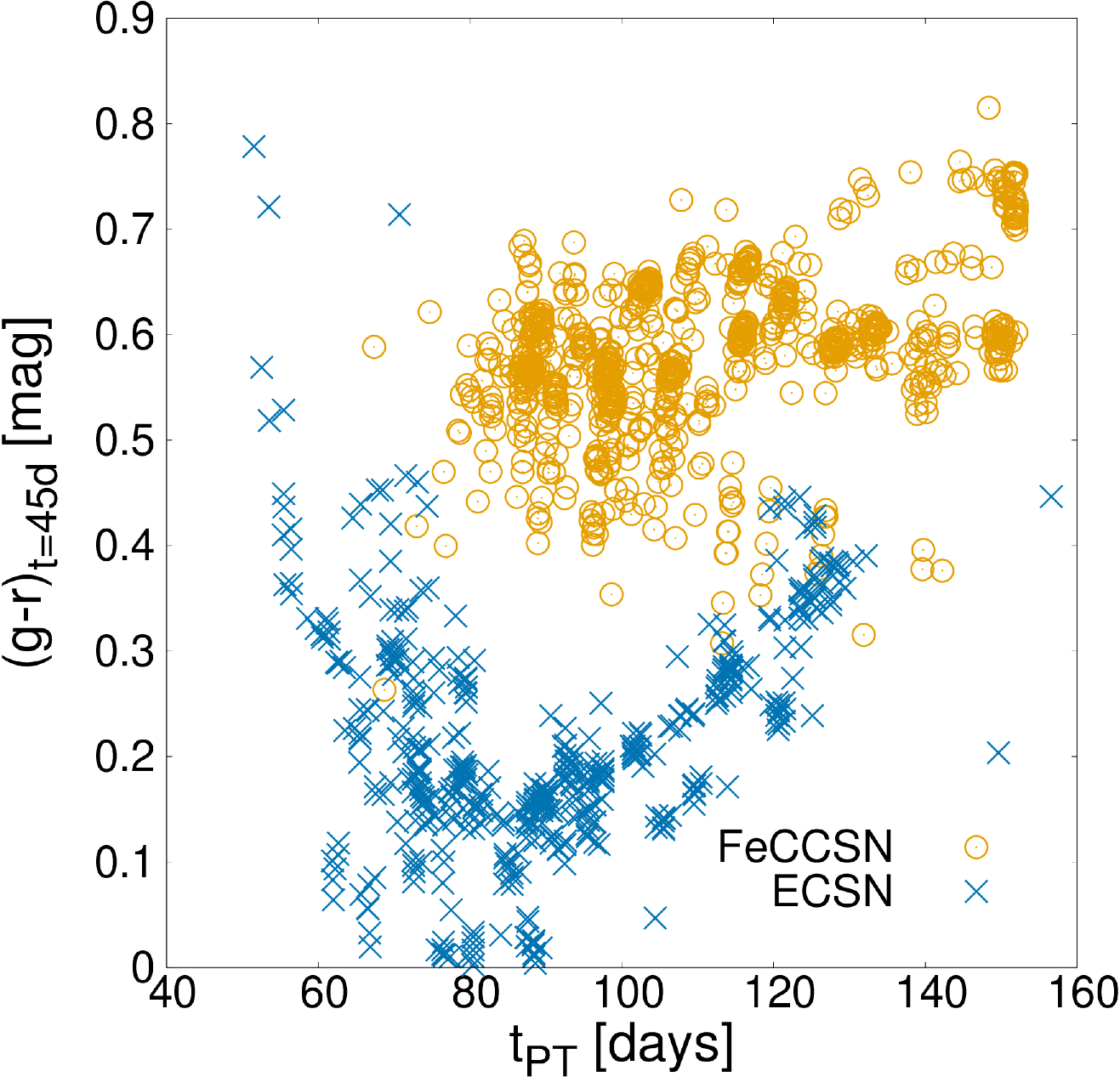}
  \end{minipage} &
  \begin{minipage}[b]{0.3\linewidth}
    \centering
    \includegraphics[width=50truemm]{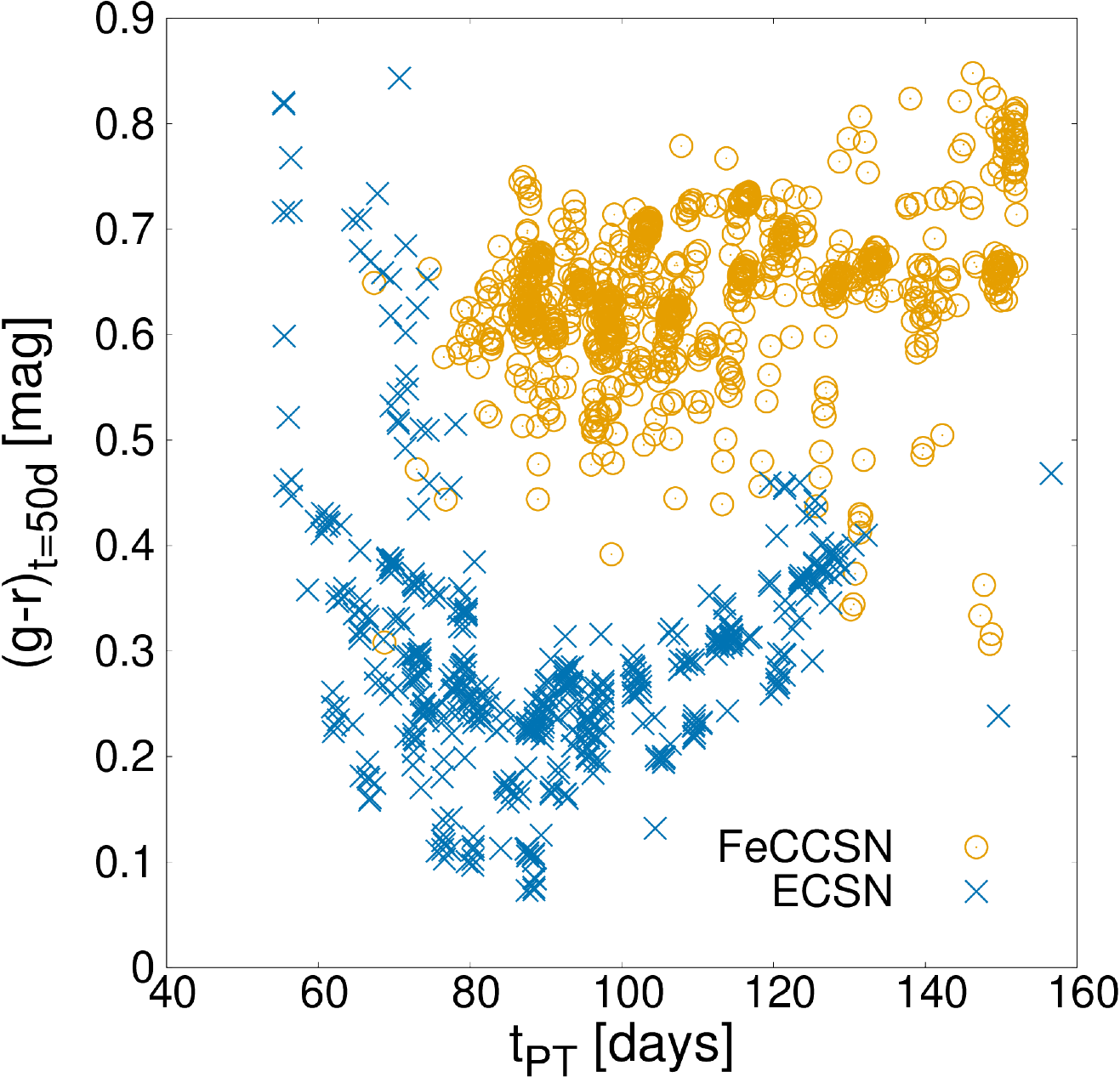}
  \end{minipage} \\
  
  \begin{minipage}[b]{0.3\linewidth}
    \centering
    \includegraphics[width=50truemm]{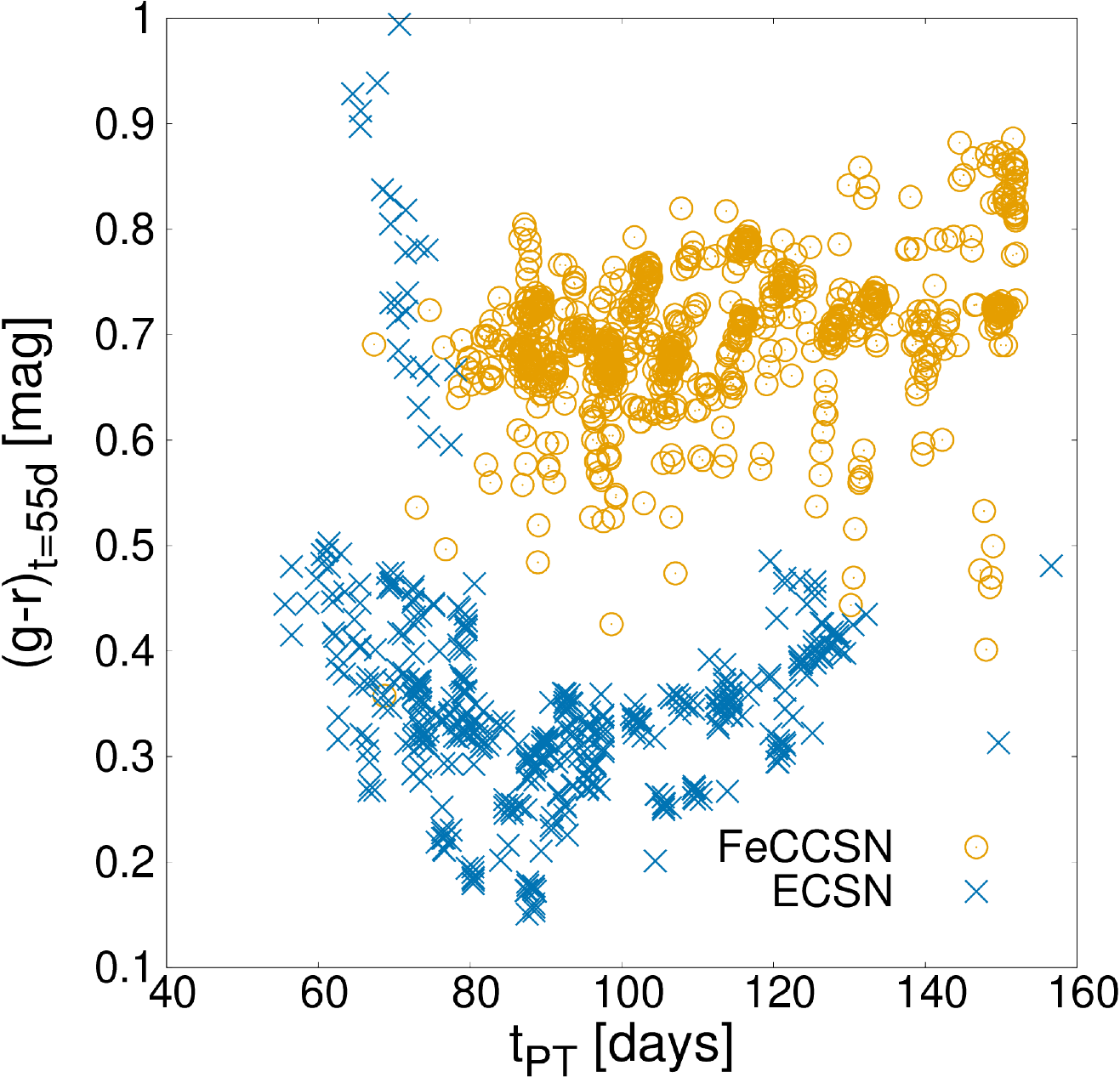}
  \end{minipage} &
  \begin{minipage}[b]{0.3\linewidth}
    \centering
    \includegraphics[width=50truemm]{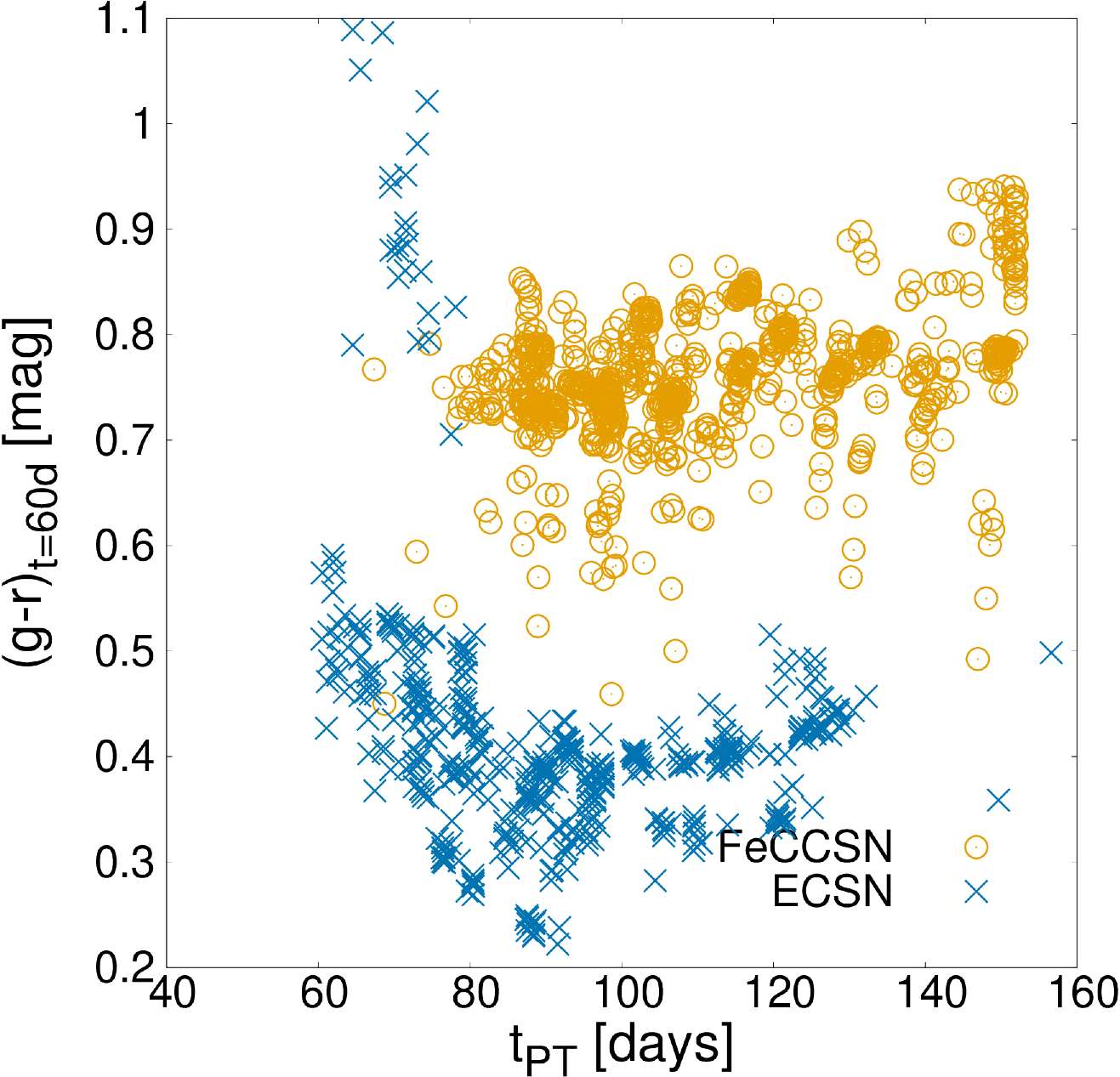}
  \end{minipage} &
  \begin{minipage}[b]{0.3\linewidth}
    \centering
    \includegraphics[width=50truemm]{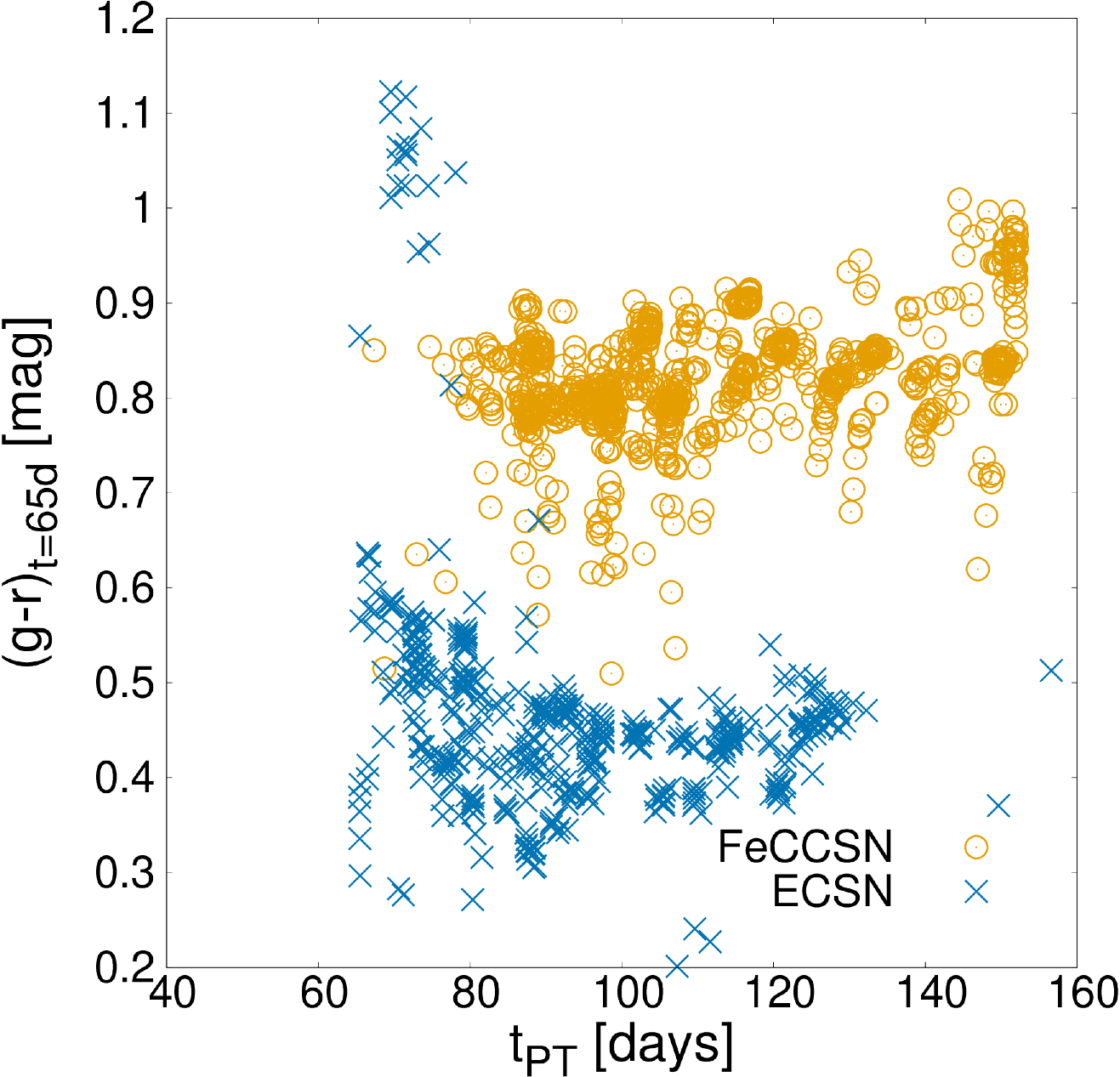}
  \end{minipage}
\end{tabular}
\caption{
ECSN indicators based on $\tPT$ and $g-r$ at 40, 45, 50, 55, 60, and 65 days.
The horizontal axis shows \(\tPT\) and the vertical axis shows $g-r$ at each epoch.
In each panel, models with earlier $\tPT$ or later CSM-interaction ends than each epoch are not shown.
\label{fig:tpt_grepoch}}
\end{figure}

\begin{figure}[ht!]
\begin{tabular}{ccc}
  \begin{minipage}[b]{0.3\linewidth}
    \centering
    \includegraphics[width=50truemm]{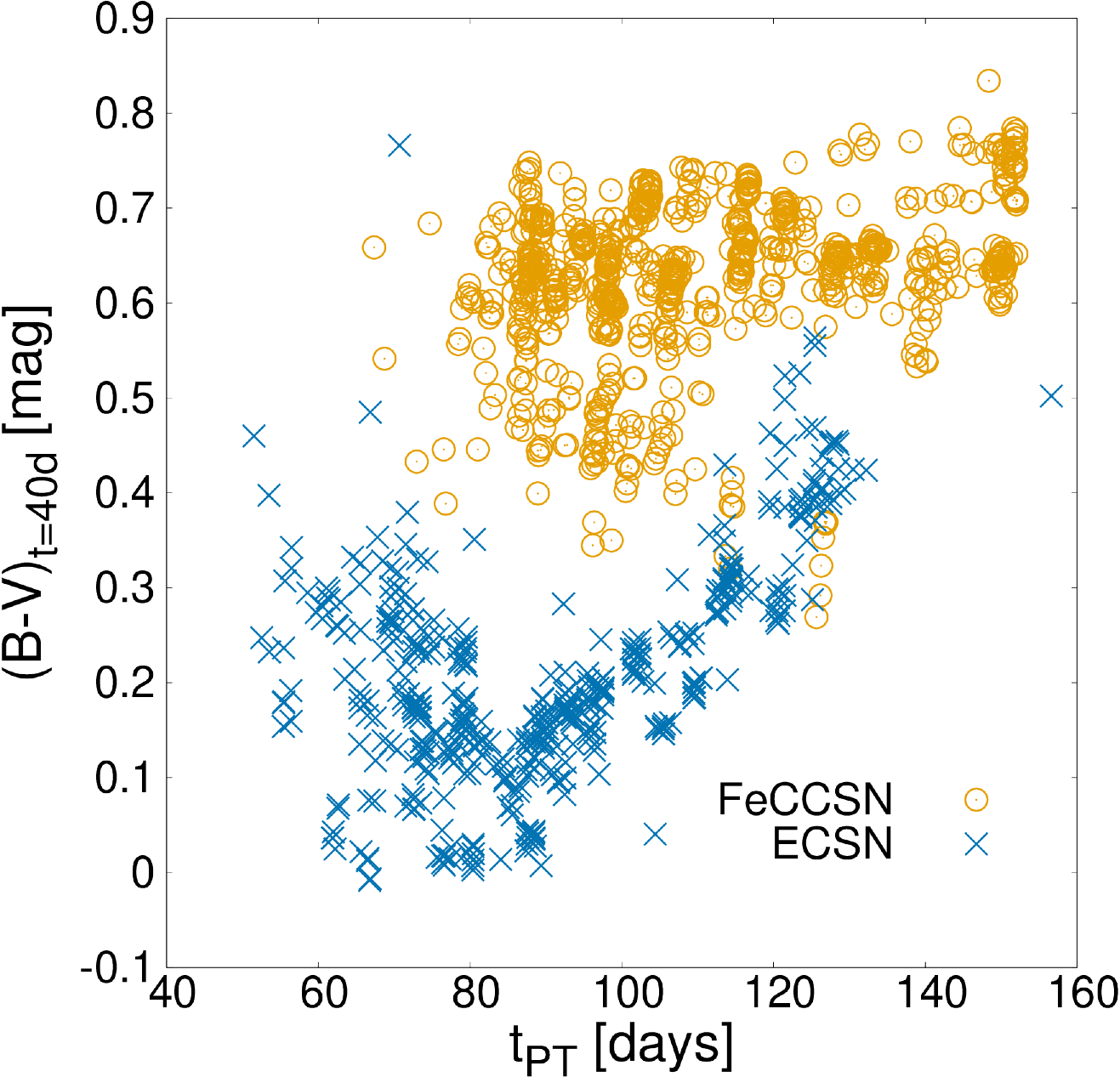}
  \end{minipage} &
  \begin{minipage}[b]{0.3\linewidth}
    \centering
    \includegraphics[width=50truemm]{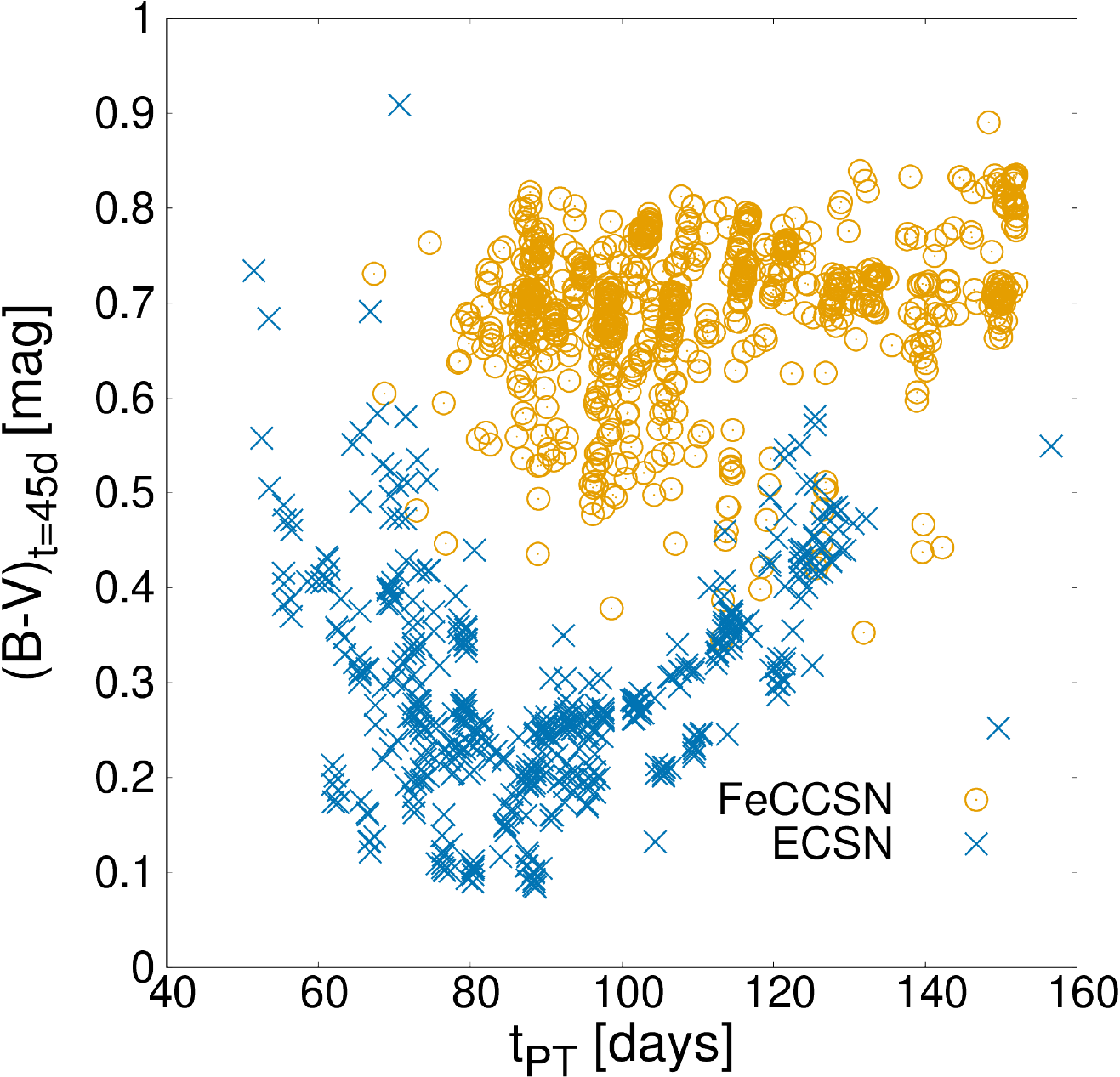}
  \end{minipage} &
  \begin{minipage}[b]{0.3\linewidth}
    \centering
    \includegraphics[width=50truemm]{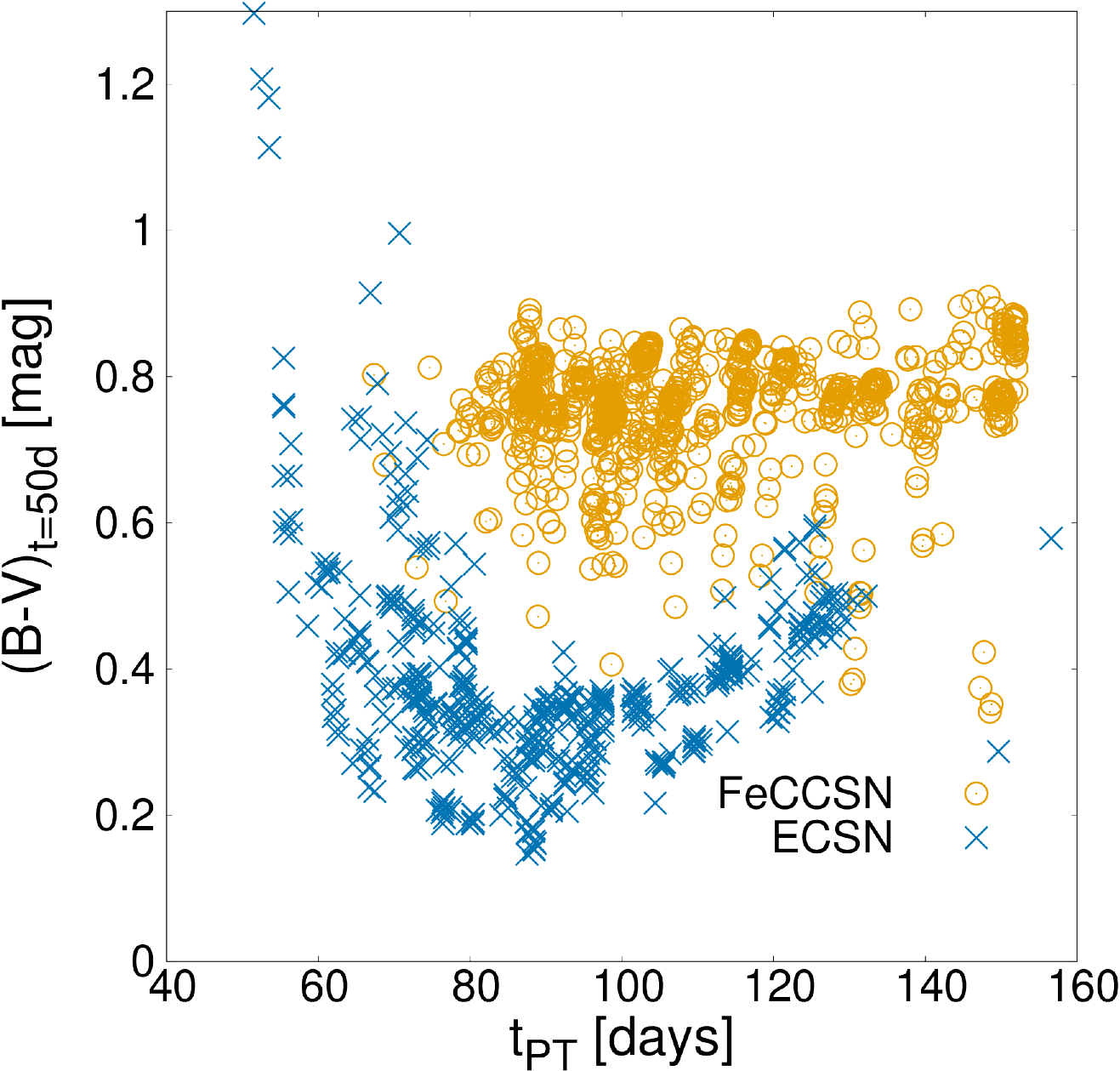}
  \end{minipage} \\
  
  \begin{minipage}[b]{0.3\linewidth}
    \centering
    \includegraphics[width=50truemm]{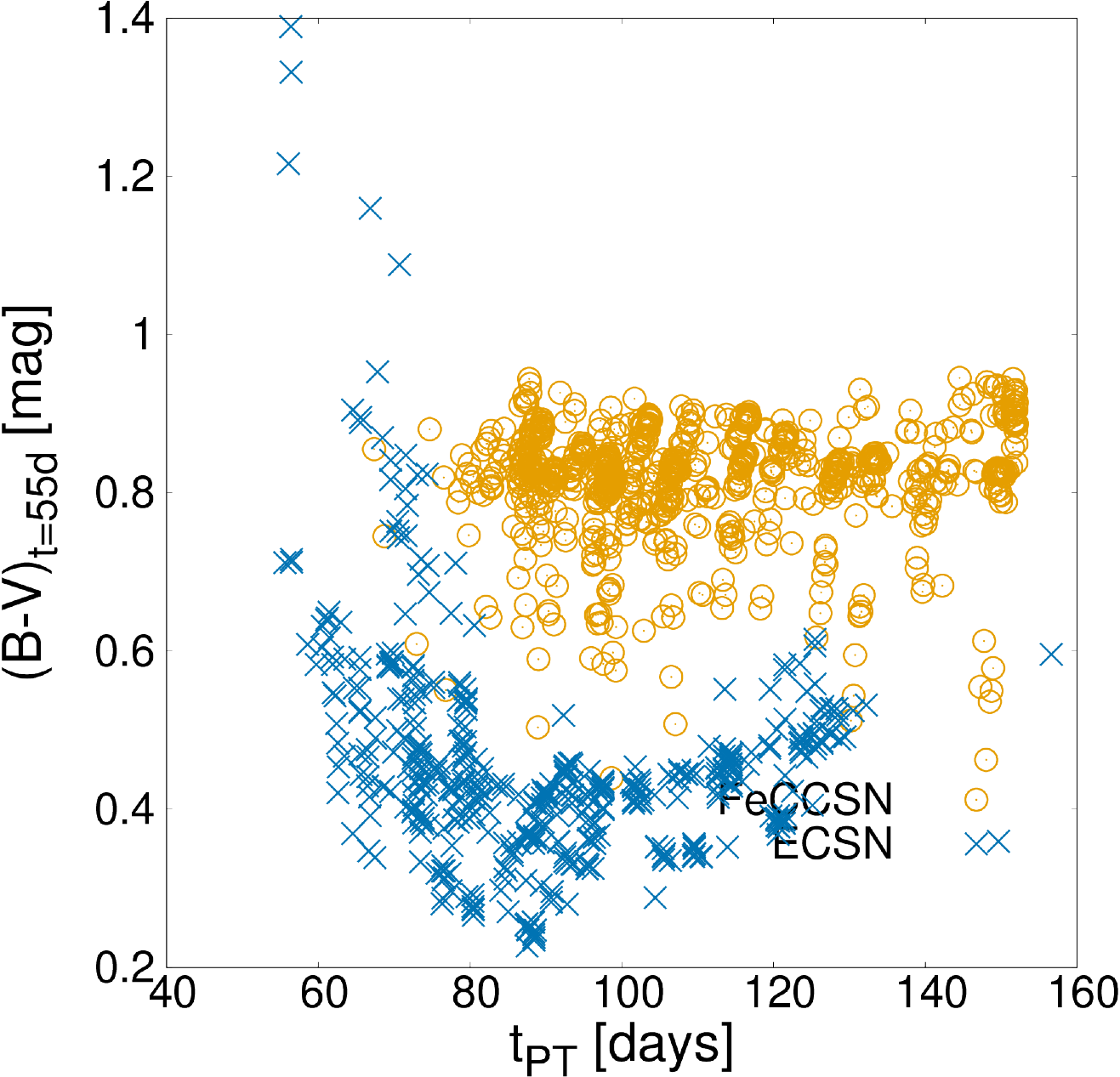}
  \end{minipage} &
  \begin{minipage}[b]{0.3\linewidth}
    \centering
    \includegraphics[width=50truemm]{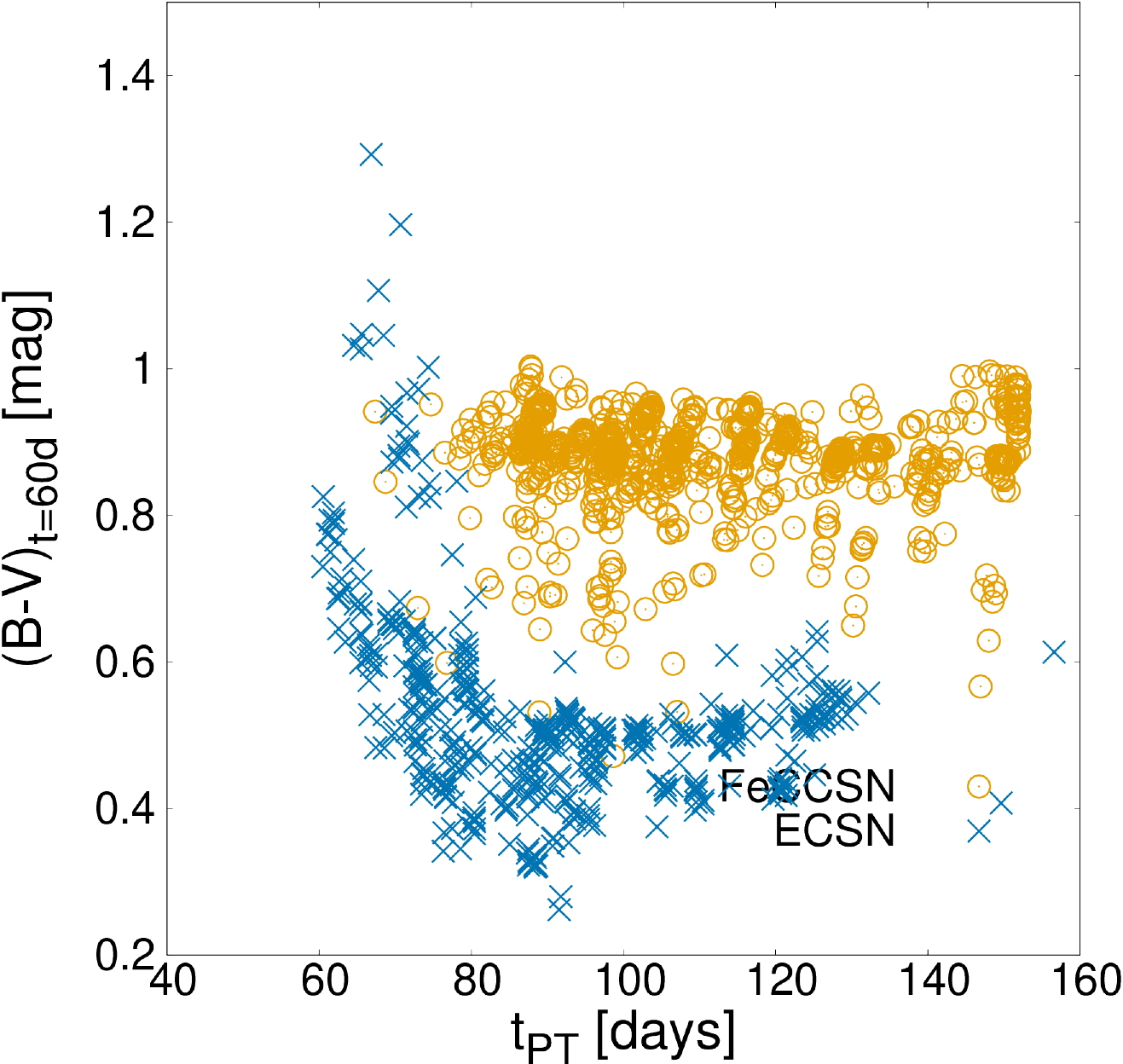}
  \end{minipage} &
  \begin{minipage}[b]{0.3\linewidth}
    \centering
    \includegraphics[width=50truemm]{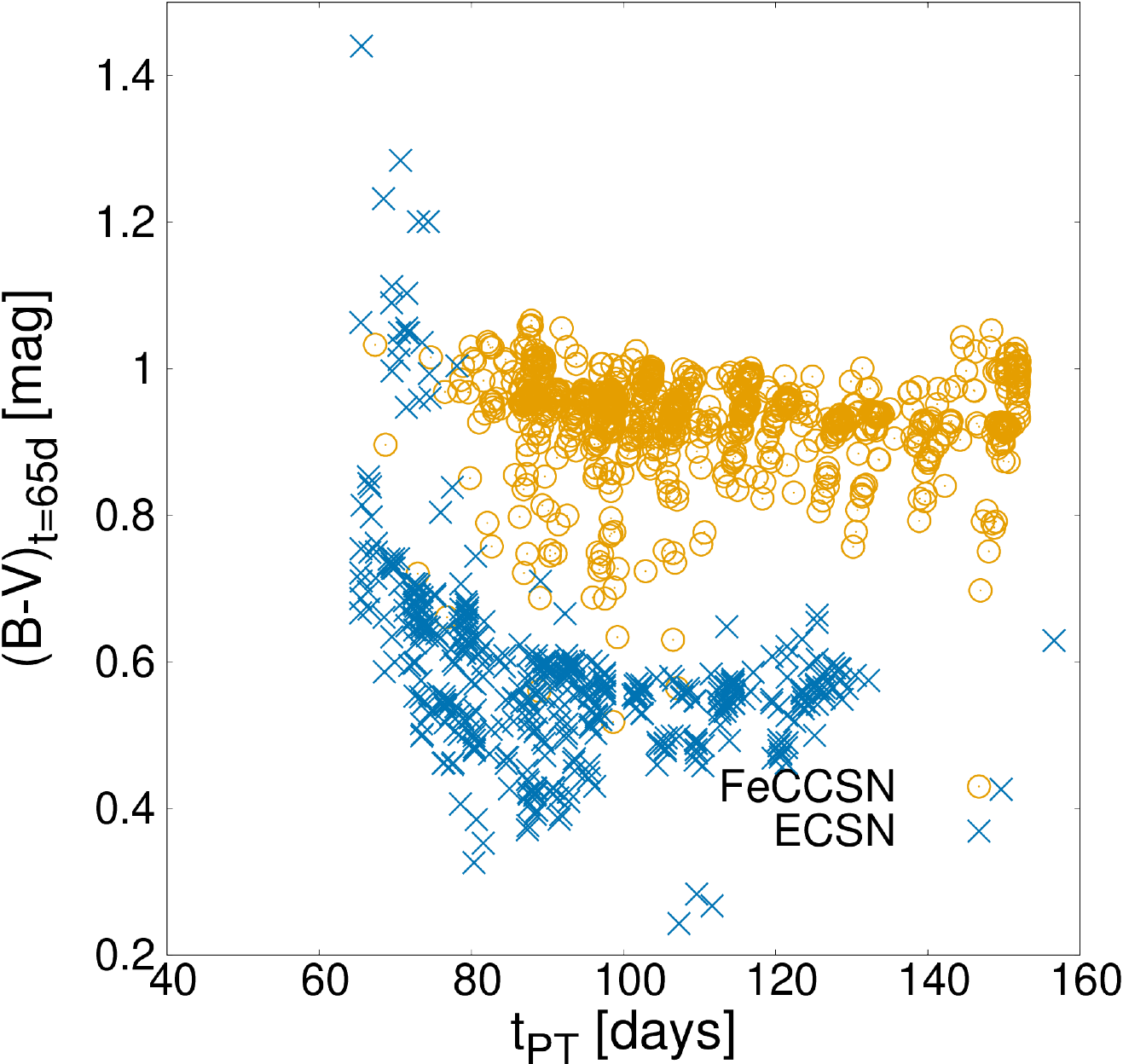}
  \end{minipage}
\end{tabular}
\caption{
Same as Figure~\ref{fig:tpt_grepoch} but for $B-V$.
\label{fig:tpt_BVepoch}}
\end{figure}

To overcome this, we adopt $g-r$ or $B-V$ at specific epochs ($t = 40,~45,~50,~55,~60,$ and $65$ days; time origin as $t_0$, defined in Section~\ref{subsec:identification}) instead of $(g-r)_{\tPT/2}$.
Figure~\ref{fig:tpt_grepoch} and \ref{fig:tpt_BVepoch} show the relation between $\tPT$ and $g-r$, and $\tPT$ and $B-V$, respectively, for each epoch.
The models with earlier plateau ends or later CSM-interaction ends than each epoch are not shown.
The colors of ECSNe and FeCCSNe are generally different, but they are similar if the adopted epoch is close to $\tPT$ (left side of each panel). This infers that earlier spectroscopy is preferred but also often suffers from CSM interaction.
Therefore, we suggest that spectroscopy around $t=50$ days is preferential if the opportunity of spectroscopic follow-ups is limited.

\section{The procedure of observations
\label{appsec:obs_procedure}}
\begin{figure}[ht!]
  \begin{minipage}[b]{0.45\linewidth}
    \centering
    \includegraphics[width=80truemm]{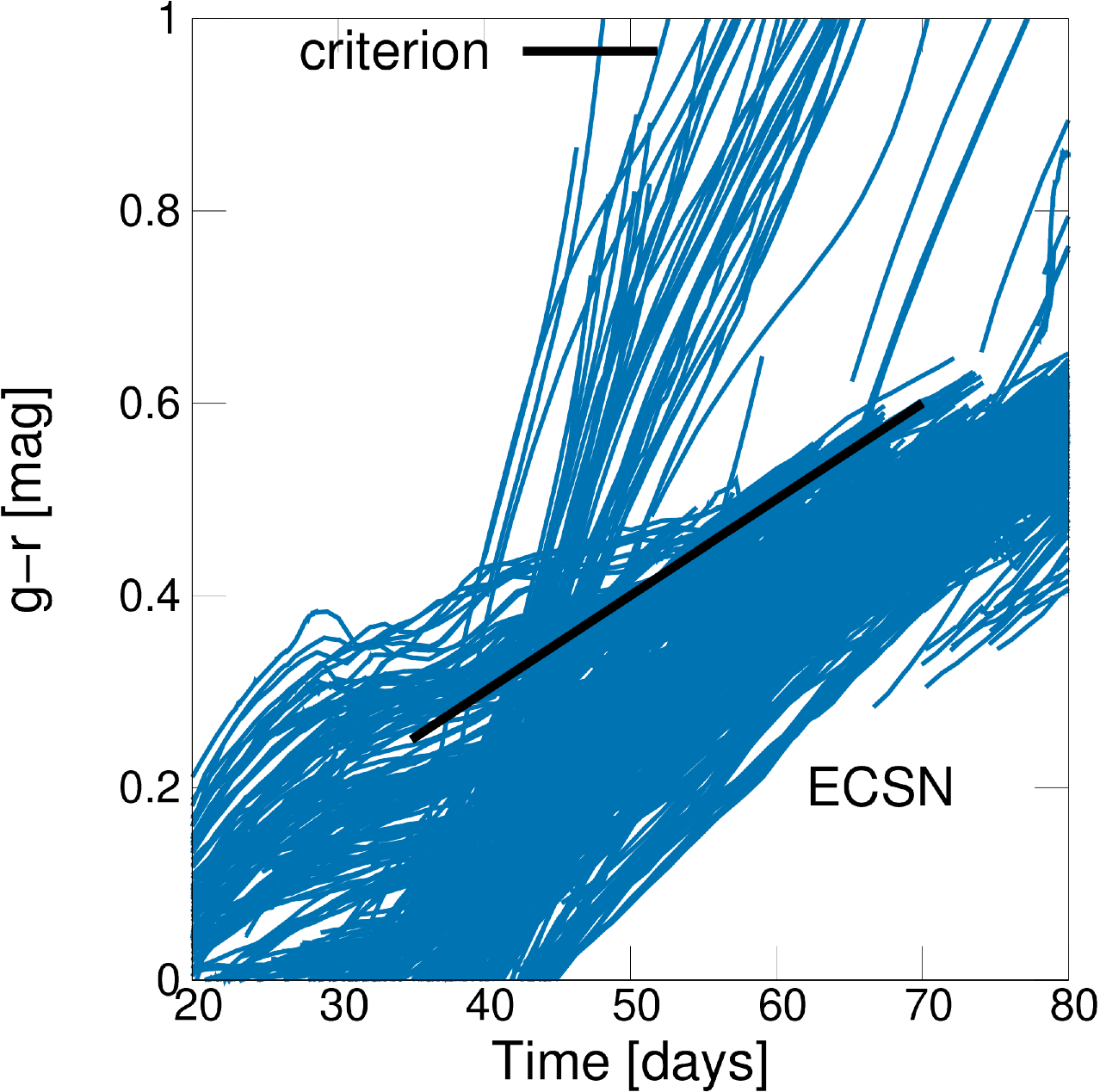}
  \end{minipage}
  \begin{minipage}[b]{0.45\linewidth}
    \centering
    \includegraphics[width=80truemm]{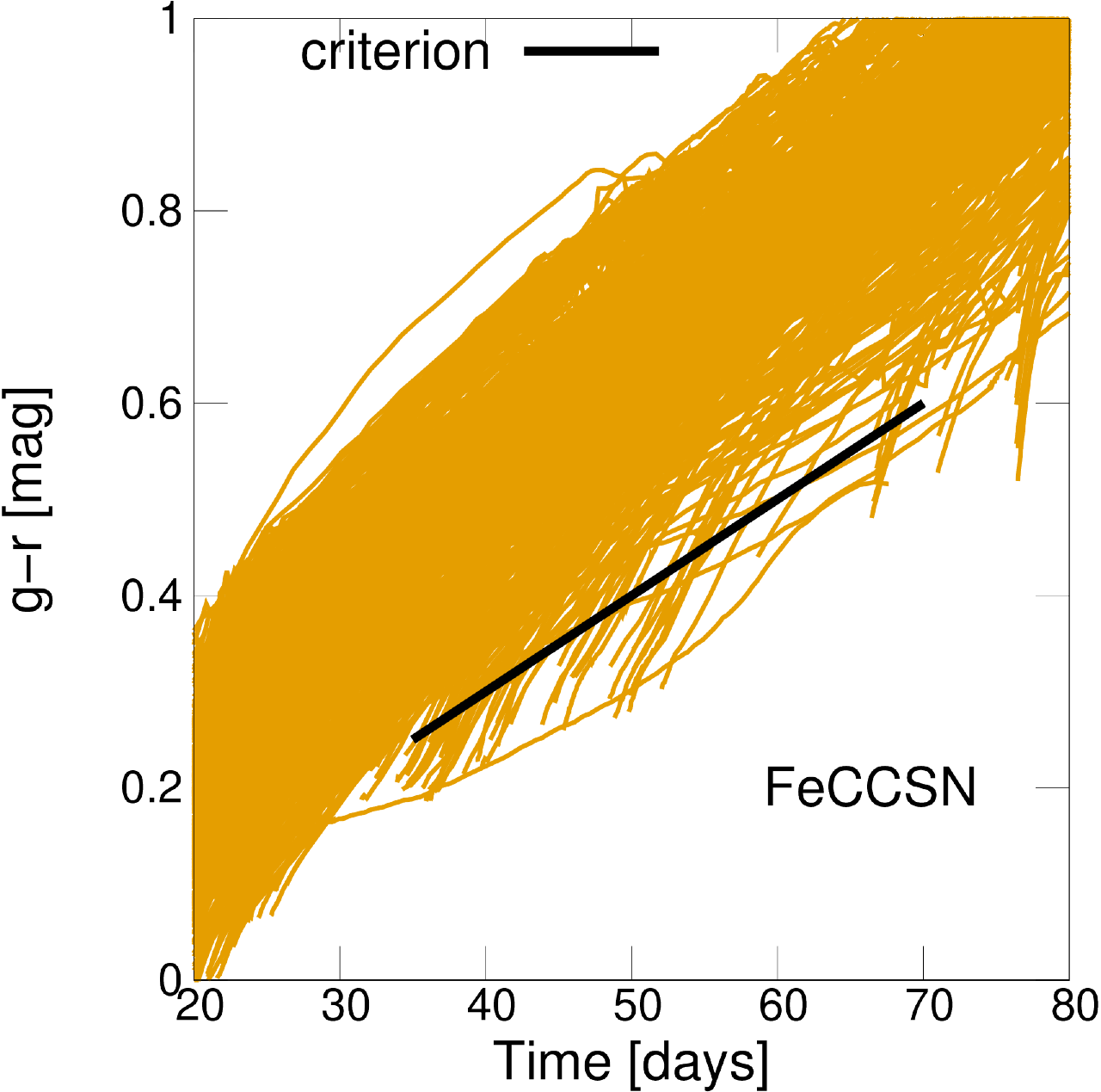}
  \end{minipage}
\caption{
A selection criterion of ECSN candidates, in which the $g-r$ evolution is adopted. 
The $g-r$ evolution of ECSNe (blue lines in the left panel) and FeCCSNe (orange lines in the right panel) are also shown though the light curves when the CSM interaction is taking place or after $\tPT$ are not shown.
If a SN~II shows bluer color than the criterion, we propose spectroscopic follow-ups to check if CSM interaction is taking place.
\label{fig:t_gr}}
\end{figure}

\begin{figure}[ht!]
  \begin{minipage}[b]{0.45\linewidth}
    \centering
    \includegraphics[width=80truemm]{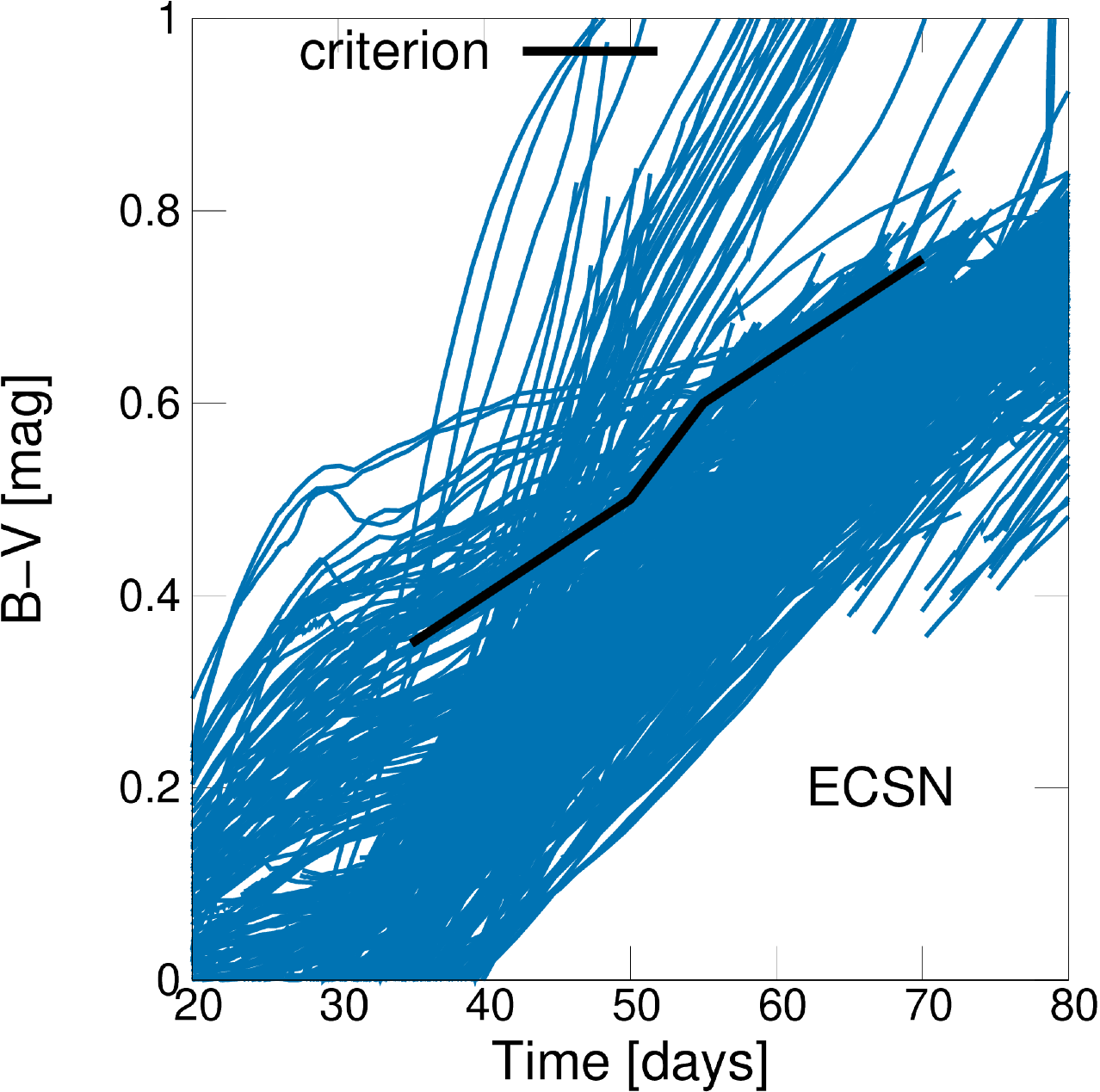}
  \end{minipage}
  \begin{minipage}[b]{0.45\linewidth}
    \centering
    \includegraphics[width=80truemm]{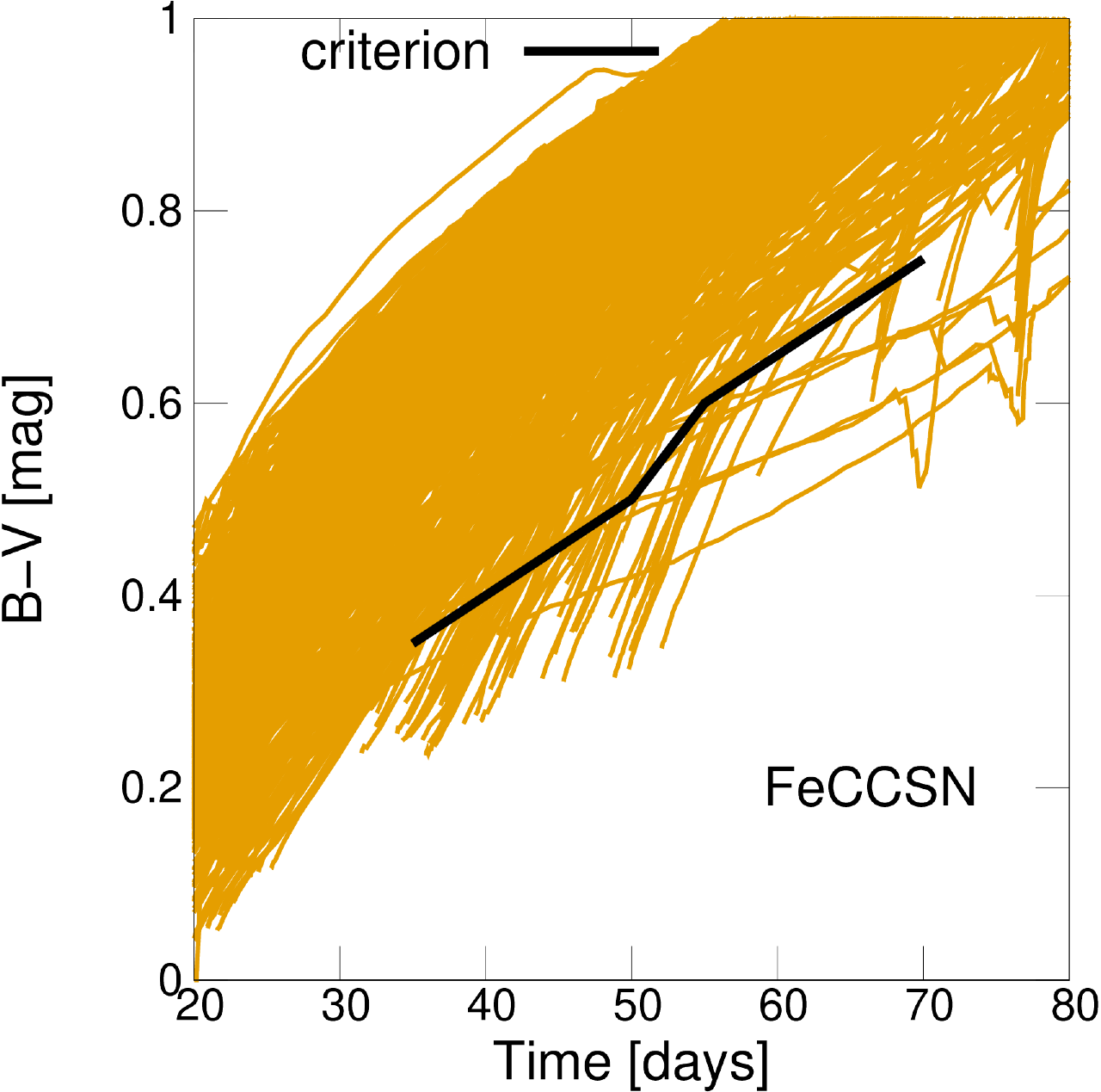}
  \end{minipage}
\caption{
Same as Figure~\ref{fig:t_gr} but $B-V$ is adopted.
\label{fig:t_BV}}
\end{figure}

We propose a procedure of observation to identify ECSNe as follows:
\begin{enumerate}
    \item 
    Explore SNe~II showing blue plateaus, which are the ECSN candidates using the selection method presented below (Figure~\ref{fig:t_gr} and \ref{fig:t_BV}), in which the evolution of $g-r$ or $B-V$ is adopted.
    \item 
    From spectroscopic observations, check whether the CSM interaction ends earlier than \(\tPT/2\).
    \item 
    If the CSM interaction ends before $\tPT/2$, identify ECSNe with equation (\ref{eq:g-r(tPT/2)threshold}) or (\ref{eq:B-V(tPT/2)threshold}). If the CSM interaction lasts beyond $\tPT/2$ but ends at later epochs in the plateau, identify ECSNe using Figure~\ref{fig:tpt_grepoch} or \ref{fig:tpt_BVepoch}.
\end{enumerate}

We present methods to select SNe~II showing blue color, which are ECSN candidates, in which the evolution of $g-r$ or $B-V$ is adopted. 
Figures~\ref{fig:t_gr} and \ref{fig:t_BV} show the criteria of ECSN candidates as well as the $g-r$ (Figure~\ref{fig:t_gr}) and $B-V$ (Figure~\ref{fig:t_BV}) evolution of ECSNe (blue lines in left panels) and FeCCSNe (orange lines in right panels) though the light curves at epochs with ongoing CSM interaction or after the plateau are not shown.
If a SN~II shows bluer light curves than the criteria, we propose spectroscopic observations to check if the CSM interaction is ongoing.

\end{document}